\documentclass{article}

\newcommand*\diff{\mathop{}\!\mathrm{d}}

\usepackage[top=0.95in, bottom=0.95in, left=0.9in, right=0.9in]{geometry}
\usepackage{setspace}

\usepackage[T1]{fontenc} 
\usepackage[round]{natbib} 
\usepackage{graphicx} 
\usepackage{color} 
\usepackage{amsmath,amssymb,theorem} 
\usepackage{listings} 
\usepackage{booktabs} 
\usepackage{rotating} 
\usepackage{hyperref} 
\usepackage{microtype} 
\usepackage{mathtools}
\usepackage{changepage}
\usepackage{bm}
\usepackage{verbatim}
\usepackage{graphicx}
\usepackage{caption}
\usepackage{subcaption}
\usepackage{natbib}
\usepackage{natbib}
\usepackage{bm}

\definecolor{deepcarmine}{rgb}{0.66, 0.13, 0.24}

\usepackage{hyperref}
\hypersetup{
	colorlinks=true,       
	linkcolor=black,          
	citecolor=deepcarmine,        
}

\DeclareCaptionLabelFormat{adja-page}{\hrulefill\\#1 #2 \emph{(previous page)}}

\begin{document}

\noindent{\LARGE \bf A geometrical analysis of global stability in trained feedback networks}
\vspace{0.5cm}

\noindent \large{{Francesca Mastrogiuseppe \textsuperscript{\bf 1,2}, Srdjan Ostojic \textsuperscript{\bf 1}}}
\vspace{0.4cm}

\normalsize{
	\noindent\textsuperscript{\sffamily{\bf 1}} Laboratoire de Neurosciences Cognitives et Computationelles, INSERM U960 and
	\\
	\textsuperscript{\sffamily{\bf 2}} Laboratoire de Physique Statistique, CNRS UMR 8550 \\
	\'Ecole Normale Sup\'erieure - PSL Research University, Paris, France
	\\}

\vspace{0.5cm}

Recurrent neural networks have been extensively studied in the context of neuroscience and machine learning due to their ability to implement complex computations. While substantial progress in designing effective learning algorithms has been achieved in the last years, a full understanding of trained recurrent networks is still lacking. Specifically, the mechanisms that allow computations to emerge from the underlying recurrent dynamics are largely unknown.
Here we focus on a simple, yet underexplored computational setup: a feedback architecture trained to associate a stationary output to a stationary input. 
As a starting point, we derive an approximate analytical description of global dynamics in trained networks which assumes uncorrelated connectivity weights in the feedback and in the random bulk. The resulting mean-field theory suggests that the task admits several classes of solutions, which imply different stability properties. Different classes are characterized in terms of the geometrical arrangement of the readout with respect to the input vectors, defined in the high-dimensional space spanned by the network population. We find that such approximate theoretical approach can be used to understand how standard training techniques implement the input-output task in finite-size feedback networks. In particular, our simplified description captures the local and the global stability properties of the target solution, and thus predicts training performance. 

\section{\textsc{Introduction}}

Over the last decades, trained networks have been used as a test-bed for understanding how complex computations can be performed by large ensembles of elementary non-linear units \citep{Rosenblatt1958, Tsoi, LeCun2015}. Trained recurrent networks, moreover, can be interpreted as models of cortical circuits  that one can  easily probe and analyze in an \emph{in-silico} setup \citep{Mante2013, Sussillo2014, Rajan2016, Wang2018, Barak2017}.
Because of the long temporal dependencies generated by the intricate connectivity, designing efficient learning algorithms for recurrent networks is, however, a difficult task.
Only recently, advances in technology and machine learning algorithms \citep{Atiya, MartensSutskever, Pascanu2013} have made possible to effectively train large recurrent networks by overcoming severe stability issues \citep{Doya, Bengio}. 

A first strategy to circumvent the problem of dealing with recurrent temporal dependencies was proposed in the work by Jaeger \citep{JaegerHaas} and Maass \citep{Maass2007}.
In these frameworks, external feedback loops were implemented to control and reshape the disordered dynamics that spontaneously appear in large random networks. In such feedback architectures, the task-specific component of the recurrent dynamics is totally specified by the feedback input from the readout unit to the random reservoir. As a result, learning can be restricted to the readout weights, which are trained by clamping the feedback input to the target of the task \citep{Jaeger, Lukosevicius}, or by fast weights modification within an online training scheme \citep{Jaeger2002, SussilloAbbott}.

From a mathematical perspective, feedback loops can be interpreted as rank-one perturbations of the originally random connectivity.
Such a low-rank perturbation can significantly modify the characteristics of the original matrix \citep{Tao2013}, and generates the articulate dynamics which are required by complex tasks like pattern generation and decision making \citep{SussilloAbbott}. In order to obtain arbitrary activity profiles, the entries of the rank-one term are often tightly fine-tuned to the entries of the original random matrix. As learning potentially relies on amplifying the correlations which exist between the readout vector and the finite-size random bulk, developing a formal understanding of the resulting network dynamics has proved to be a difficult theoretical task \citep{SussilloBarak, MassarMassar, RivkindBarak}.

In this work, we focus on a simple, yet not fully explored scenario: the feedback network is trained to produce a stationary output signal in response to a constant pattern of external inputs \citep{RivkindBarak}.
To characterize the problem from a theoretical perspective, we focus on a restricted class of simple training solutions for which a complete analytical description can be developed. 
These training solutions correspond to readouts which are strongly aligned with the external and feedback input weights, but are uncorrelated with respect to the random synaptic weights within the reservoir. For such readouts, we derive a mean-field theory which captures the network behaviour in terms of a few representative macroscopic variables \citep{Sompolinsky1988, MastrogiuseppeOstojic2} and allows us to isolate the relevant geometrical components of the readout vector which dominantly contribute to the final network dynamics. In particular, we find that different combinations of these components can be mapped into different output states of the final feedback network.

In a second step, we consider the more general training solutions which are found by standard learning techniques, such as least-squares (LS) \citep{Jaeger, Lukosevicius} and recursive least-squares (RLS) algorithms \citep{SussilloAbbott, LajeBuonomano}. We use our mean-field framework to derive approximate analytical descriptions of such solutions, where the full trained readout vector is replaced by its projection on the hyperplane spanned by the external and feedback input vectors. We show that such simplified portraits predict with good precision the output states that are displayed by trained networks, together with their stability: our simplified description completely neglects the correlations existing between the readout vector and the random bulk, but still accurately predicts how the network phase space is reshaped by training. Specifically, we show that learning through the rank-one feedback can induce severe dynamical instabilities even in an extremely simple task. We use our theoretical framework to unveil and compare the strategies that the LS and RLS algorithms adopt to solve the task, and to clarify the reasons of their failures and successes.

\section{\textsc{The setup}}

We consider a rate-based network model, where every unit $i$ is characterized by a continuous activation variable $x_i$ and its non-linear transform $\phi(x_i)$ \citep{Sompolinsky1988}. In line with many classical studies \citep{Jaeger, SussilloAbbott, LajeBuonomano, RivkindBarak}, we chose $\phi(x)=\tanh(x)$. The case of rate networks characterized by a threshold-linear activation function is discussed in \emph{Appendix C}.

To begin with, we consider a random reservoir network driven by an external input consisting of an $N-$dimensional stationary pattern $\mathbf{I}$. Its dynamics read:
\begin{equation}\label{eq:dyn_nofeedback}
\dot{x_i}(t) = -x_i(t) + g\sum_{i=1}^N \chi_{ij} \phi(x_j(t)) + I_i,
\end{equation}
where we have rescaled time to set the time constant to unity. The parameter $g$ quantifies the strength of the random connectivity $\bm{\chi}$, whose entries are generated from a centered Gaussian distribution of variance $1/N$ \citep{Sompolinsky1988}. 
The network output is defined at the level of a linear readout:
\begin{equation}\label{eq:readout}
	z(t) =\sum_{i=1}^N n_i \phi(x_i(t)),
\end{equation}
where vector $\mathbf{n}$ sets the output readout direction. 

A task specifies how the network needs to map the external input $\mathbf{I}$ into the output signal $z(t)$.
The task is solved when an appropriate readout vector $\mathbf{n}$ has been learned, such that the readout $z$ correctly matches the target. 
Here we consider a specific fixed point task, in which the network has to hold an arbitrary stationary output signal: $z(t) = A$.

The random reservoir in Eq.~\ref{eq:dyn_nofeedback} admits a unique input-driven stable state \citep{Kadmon2015, MastrogiuseppeOstojic}, where single units activity is a combination of the feedforward input and the recurrent dynamics generated by random synapses. For reasons that will become clear in a moment, we refer to this attractor as the \emph{feedback-free} activity state. 
Once the network activity has relaxed into such a state, the task can be solved by tuning the readout weights $n_i$ until the network output matches the target $A$. This training protocol results into a \emph{feedback-free} implementation of the task, for which the dynamics of the reservoir is independent of the choice of the readout weights. 

While they are suitable for implementing many simple input-output associations, feedback-free reservoirs have a limited range of applications. As pointed out in a series of seminal studies \citep{Jaeger, Maass2007, SussilloAbbott}, more flexible architectures can be obtained by using the readout signal $z(t)$ as an additional external input, which acts as an external feedback to the random reservoir (Fig.~\ref{fig:general} \textsf{a}). The dynamics of the network obey:
\begin{equation}\label{eq:full_dynamics}
	\dot{x_i}(t) = -x_i(t) + g\sum_{i=1}^N \chi_{ij} \phi(x_j(t)) + m_i z(t) + I_i,
\end{equation}
where the $N$-dimensional vector $\mathbf{m}$ defines the input direction of the feedback signal. In standard training procedures \citep{Jaeger, SussilloAbbott}, vector $\mathbf{m}$ is considered to be fixed across learning, and is generated with random entries. 

The feedback loop from the readout to the reservoir introduces a novel, one-dimensional element of recurrent dynamics in the network.
Substituting Eq.~\ref{eq:readout} into Eq.~\ref{eq:full_dynamics} allows to verify that the final feedback architecture can be described by an equivalent connectivity matrix, which contains the original random bulk together with a rank-one matrix specified by the feedback vectors:
\begin{equation}\label{eq:full_dynamics_rankone}
\dot{x_i}(t) = -x_i(t) + \sum_{i=1}^N \left( g\chi_{ij} + m_i n_j \right) \phi(x_j(t)) + I_i.
\end{equation}
As observed in a variety of studies \citep{Jaeger, Lukosevicius}, such a low-rank matrix perturbation can deeply reshape the dynamical landscape of the reservoir network. The number of activity states and their stability properties, which directly affect training performance in practical applications, strongly depend on the choice of the readout vector $\mathbf{n}$. Building an exact causal relationship between the different readout solutions and the corresponding emerging dynamics is, however, a non-trivial theoretical problem \citep{MassarMassar, RivkindBarak, Landau}.

In this work, we characterize the dynamics emerging in feedback networks (Eq.~\ref{eq:full_dynamics}) trained to solve the fixed point task. 
The input vectors $\mathbf{m}$ and $\mathbf{I}$ determine the fixed geometry of the task by specifying two preferred directions in the $N$-dimensional phase space spanned by the reservoir population. 
In our framework, vectors $\mathbf{m}$ and $\mathbf{I}$ are generated as random patterns of variance $\sigma_m^2$ and $\sigma_I^2$. Their entries are extracted from joint Gaussian distributions of null mean. The two vectors can thus be constructed as:
 \begin{equation}\label{eq:mI}
 \begin{split}
 &\mathbf{m} = \sigma_m \left( \rho  \bm{\xi}  + \sqrt{1 - \rho^2} \: \bm{\eta}_m \right) \\
 &\mathbf{I} = \sigma_I \left( \rho  \bm{\xi}  + \sqrt{1 - \rho^2} \: \bm{\eta}_I \right) \\
 \end{split}
 \end{equation}
 where $\bm{\xi}$, $\bm{\eta}_m$ and $\bm{\eta}_I$ are standard Gaussian vectors.
The parameter $\rho$ quantifies the strength of correlations between the input vectors $\mathbf{m}$ and $\mathbf{I}$, defined onto the overlap direction $\bm{\xi}$ (alternatively, the overlap direction $\bm{\xi}$ can be taken to be parallel to the unitary vector $\mathbf{u} = (1, ..., 1)$, resulting in input vectors $\mathbf{m}$ and $\mathbf{I}$ characterized by random and uncorrelated entries of non-zero mean). From this setup, the feedback-free implementations can be retrieved as a special case by setting the feedback inputs to zero through $\sigma_m=0$.

\section{\textsc{General classes of solutions}}
As a first step, we approach the problem from a theoretical perspective. We analyze the space of all the possible task solutions by focusing on the limit of  a large random bulk ($N\rightarrow\infty$).

The specific task we consider requires a constant readout signal, which is trivially obtained in stationary network states. Equilibrium solutions, which are obtained by setting $\dot{x}_i(t)=0$ in Eq.~\ref{eq:full_dynamics}, can be written as:
\begin{equation}\label{eq:xi}
x_i = m_i z + I_i + \delta_i,
\end{equation}
where the term $\delta_i$ corresponds to self-consistent recurrent input generated by the random part of the connectivity:
\begin{equation}
\delta_i = g\sum_{j=1}^N \chi_{ij} \phi(x_j(t)).
\end{equation}
The value of the readout $z$ can thus be re-expressed as:
\begin{equation}\label{eq:kappa}
\begin{split}
z & = \sum_{i=1}^N n_i \phi(m_i z + I_i + \delta_i)\\
& = N \langle n_i \: \phi (m_i z + I_i + \delta_i) \rangle,
\end{split}
\end{equation}
where the notation $\langle . \rangle$ indicates an average over the population of units. 

For a broad range of $\phi(x)$ functions, the output $z$ is thus determined by the correlations between the readout $\mathbf{n}$ and the vectors defining the direction of the network activity: $\mathbf{m}$, $ \mathbf{I}$ and $\bm{\delta}$.
While vectors $\mathbf{m}$ and  $\mathbf{I}$ are considered to be fixed, the direction of the residual input $\bm{\delta}$ varies from one realization to the other of the random connectivity $\bm{\chi}$. For a given matrix $\bm{\chi}$, furthermore, the direction of vector $\bm{\delta}$ varies in the different dynamical states admitted by the network. For these reasons, the residual input $\bm{\delta}$ represents the hardest term to evaluate and control.

One approach consists in fixing the random connectivity matrix and computing $\bm{\delta}$  in the \emph{open-loop} configuration, obtained by clamping the readout signal to the target: $z(t)=A$ \citep{Jaeger, RivkindBarak}. 
Such a value for $\bm{\delta}$, however, holds only locally -- that is, in the vicinity of the target fixed point. As a consequence, this approach does not allow us to control the full, global dynamics that the readout $\mathbf{n}$ imposes on the final feedback network. For example, it does not allow to predict whether additional, spurious attractors are created together with the desired output state.

A different approach is used in classical mean-field studies of purely random networks \citep{Sompolinsky1988, Molgedey, Cessac94, Schuecker2016, Marti2018}. In that case, the recurrent input $\bm{\delta}$ is approximated by a vector of uncorrelated white noise of self-consistent amplitude. Such an approximation is justified by the disorder in the random connectivity, which effectively decorrelates the input to different units.
This procedure washes out any dependence on  the specific instantiation of the random matrix $\bm{\chi}$, so that classical mean-field theories cannot be easily used to describe solutions where the readout vector $\mathbf{n}$ is tightly correlated with $\bm{\delta}$.

For the simple task we consider here, one can focus on the restricted class of solutions where the readout $\mathbf{n}$ is only correlated with the input vectors $\mathbf{m}$ and $ \mathbf{I}$. We show in the next paragraph that the classical mean-field approaches can be easily used in this case to map any readout vector into a full description of the network  output states. As they do not rely on correlations with the random bulk $\bm{\chi}$, readout solutions in this class generate an output which is robust with respect to changes in the random part of the connectivity. 

From Eq.~\ref{eq:mI}, a minimal and general solution in this class can be written as:
\begin{equation}\label{eq:sol}
\mathbf{n} = \frac{c}{N} \left(p \bm{\xi} + p_m \bm{\eta}_m + p_I \bm{\eta}_I\right).
\end{equation}
The geometry of the readout vector is determined by the relative weights of the three coefficients $(p, p_m, p_I)$, which quantify how much the readout direction is distributed on the three axes defined by the non-trained part of the network architecture.

When the readout vector is aligned with one of the three axes $\bm{\xi}$, $\bm{\eta}_m$ or $\bm{\eta}_I$, the population average in Eq.~\ref{eq:kappa} takes finite $\mathcal{O}(1)$ values (see below). As a consequence, training solutions in the form of Eq.~\ref{eq:sol} need to scale as the inverse of the network size ($n_i \sim \mathcal{O}(1/N)$) in order to prevent diverging activity and outputs. 
The weights $(p, p_m, p_I)$ are furthermore normalized through the constant $c$,  which fixes the output $z$ to the exact target value specified by the task. 


In the remaining of this section, we characterize analytically the dynamics emerging in trained networks where the readout vector belongs to the restricted class defined by Eq.~\ref{eq:sol}. To this end, we introduce the key equations of the mean-field description and we analyze in detail three simple geometries $(p, p_m, p_I)$ which satisfy the task through rather different output dynamics. In Sections 4 and 5, we exploit our simplified mean-field framework to analyze the more general readout solutions which are generated through learning by standard training algorithms.

\subsection{Mean-field description}

For every fixed readout vector in the form of Eq.~\ref{eq:sol}, we follow \citet{MastrogiuseppeOstojic2} to derive a macroscopic, effective description of the  network activity. As in standard mean-field theories for random recurrent networks \citep{Sompolinsky1988, Rajan2010, Kadmon2015}, we approximate the equilibrium activation vector $\mathbf{x}$ by the statistical distribution of its elements. In particular, every element $x_i$ can be thought as extracted from a Gaussian distribution of variance $\Delta \equiv \langle [x_i^2]\rangle - \langle[x_i]\rangle^2$, where $\langle.\rangle$ indicates an average over the population and $[.]$ an average across different realizations of the random bulk. 

From here on, we consider the overlap direction $\bm{\xi}$ to be generated as a standard Gaussian vector. In this case, the mean of the distribution of $\mathbf{x}$ vanishes, as $\langle m_i \rangle = \langle I_i \rangle = [\chi_{ij}] =0$. As shown in \citet{MastrogiuseppeOstojic2}, if the overlap direction $\bm{\xi}$ coincides instead with the unitary vector, the mean-field equations take a slightly different form, which includes finite mean values but result in qualitatively identical results.

In order to derive a self-consistent expression for $\Delta$, we first consider the average over the bulk connectivity $\bm{\chi}$. 
By direct averaging of Eq.~\ref{eq:xi} we get:
\begin{equation}\label{eq:mf_i}
\begin{split}
&\mu_i = [x_i] = m_i [z] + I_i\\
&\Delta_i = [x_i^2] - [x_i]^2= m_i^2 \left([z^2]-[z]^2\right) + [\delta_i^2]. 
\end{split}
\end{equation}

As in standard mean-field derivations \citep{Sompolinsky1988}, we have $[x_ix_j]=[x_i][x_j]$ and:
\begin{equation}
[\delta_i^2] = g^2 \sum_{j=1}^N \sum_{k=1}^N [\chi_{ij}\chi_{ik}] [\phi(x_j)\phi(x_k)]  = g^2 \sum_{j=1}^N [\chi_{ij}^2][\phi(x_j)^2] = g^2 \langle [\phi(x_i) ^ 2]\rangle,
\end{equation}
since $[\chi_{ij}^2]=1/N$. In the thermodynamic limit, furthermore, the variance of $z$ vanishes, as $n_i$ is characterized by a weaker scaling with $N$ (from Eq.~\ref{eq:sol}, $n_i^2 \sim \mathcal{O}(1/N^2)$). The readout signal $z$ is thus self-averaging in large networks. For this reason, we drop the brackets in $[z]$, and we compute the readout signal self-consistently through:
\begin{equation}\label{eq:kappa_mf}
z  = c \:\langle \left\{ p \xi_i + p_m \eta_{mi} + p_I \eta_{Ii} \right\} [\phi(x_i)] \rangle.
\end{equation}
Note that Eq.~\ref{eq:mf_i} allows to effectively replace the recurrent input $\bm{\delta}$ with white noise of self-consistent amplitude:
\begin{equation}\label{eq:xi_approx}
x_i = \mu_i + \delta_i \sim \mu_i + \sqrt{\Delta_i} w_i
\end{equation}
where $\Delta_i = g^2 \langle [\phi(x_i) ^ 2]\rangle$ and $w_i$ is a standard Gaussian variable.

By averaging again across units, we find that the population distribution of the equilibria $x_i$ obeys the following statistics:
\begin{equation}\label{eq:mf}
\begin{split}
& \mu = \langle [x_i]  \rangle = 0 \\
& \Delta = \langle[x_i^2]\rangle - \langle[x_i]\rangle^2 = g^2 \langle [\phi(x_i) ^ 2]\rangle +\sigma_m^2 z^2 + 2\sigma_{mI} z + \sigma_I^2,
\end{split}
\end{equation}
where $\sigma_{mI} = \langle m_i I_i \rangle = \rho^2 \sigma_m\sigma_I$.

In order to obtain a closed form for our equations, we derive a self-consistent expression for the average quantities $z$ and $\langle [\phi(x_i) ^ 2]\rangle$. As in standard mean-field theories, we re-express averages through Gaussian integrals:
\begin{equation}\label{eq:gauss_int}
\langle [\phi(x_i) ^ 2]\rangle = \int \mathcal{D}w \: \phi^2 (\sqrt{\Delta} w),
\end{equation}
where we used the short-hand notation $\mathcal{D}w = \diff w e^{-\frac{w^2}{2}}/\sqrt{2\pi} $. By combining Eqs.~\ref{eq:mI} and \ref{eq:kappa_mf}, and by integrating by parts  over $\xi_i$, $\eta_{mi}$ and $\eta_{Ii}$, we furthermore get \citep{MastrogiuseppeOstojic2}:
\begin{equation}
\begin{split}\label{eq:kappa_fin}
z = c \: \left\{ p (\sigma_m \rho z + \sigma_I \rho) + p_m \sigma_m \sqrt{1-\rho^2} z + p_I \sigma_I \sqrt{1-\rho^2}\right\} \langle [\phi'(x_i)]\rangle,
\end{split}
\end{equation}
where the average $\langle [\phi'(x_i)]\rangle$ can be evaluated similarly to Eq.~\ref{eq:gauss_int}. For any fixed set of readout weights $(p,p_m,p_I)$, Eqs.~\ref{eq:mf} and \ref{eq:kappa_fin} can be solved together through standard numerical techniques to return the values of $\Delta$ and $z$ which characterize the different output fixed points.

To conclude, we observe that the equations above  only apply to networks which satisfy the task through stationary output states. In general, stationary outputs do not require stationary reservoir activity. For example, solutions in the form of Eq.~\ref{eq:sol} can be used to solve the task even when activity in the bulk is chaotic, but fluctuations cancel out at the level of the readout \citep{MastrogiuseppeOstojic2}. In that case, the mean-field description above can easily be extended to take temporal fluctuations into account \citep{Sompolinsky1988}.
Standard training techniques, however, typically fail to converge when the target of the task is not strong enough to suppress chaotic activity in the bulk \citep{RivkindBarak}. For this reason, here and in the following, we concentrate on fixed point dynamics.

\subsection{Determining task solutions}

For every fixed set of weights $(p, p_m, p_I)$, first the normalization parameter $c$ needs to be determined. We fix $c$ by requiring that, in the final network output, the readout value exactly matches the target $A$. To this end, we consider the {open-loop} formulation of Eqs.~\ref{eq:mf} and \ref{eq:kappa_fin}, obtained by setting $z=A$, and we solve the system for $c$ and $\Delta^{ol}$. This gives:
\begin{equation}\label{eq:c}
c = \frac{A}{\left\{ p (\sigma_m \rho A + \sigma_I \rho) + p_m \sigma_m \sqrt{1-\rho^2} A + p_I \sigma_I \sqrt{1-\rho^2}\right\} \langle [\phi'(x_i^{ol})]\rangle}
\end{equation}
where $\langle [\phi'(x_i^{ol})]\rangle$ is computed as Gaussian integral over a distribution of variance $\Delta^{ol}$ (see Eq.~\ref{eq:gauss_int}).

\begin{figure}

		\centering
		\includegraphics{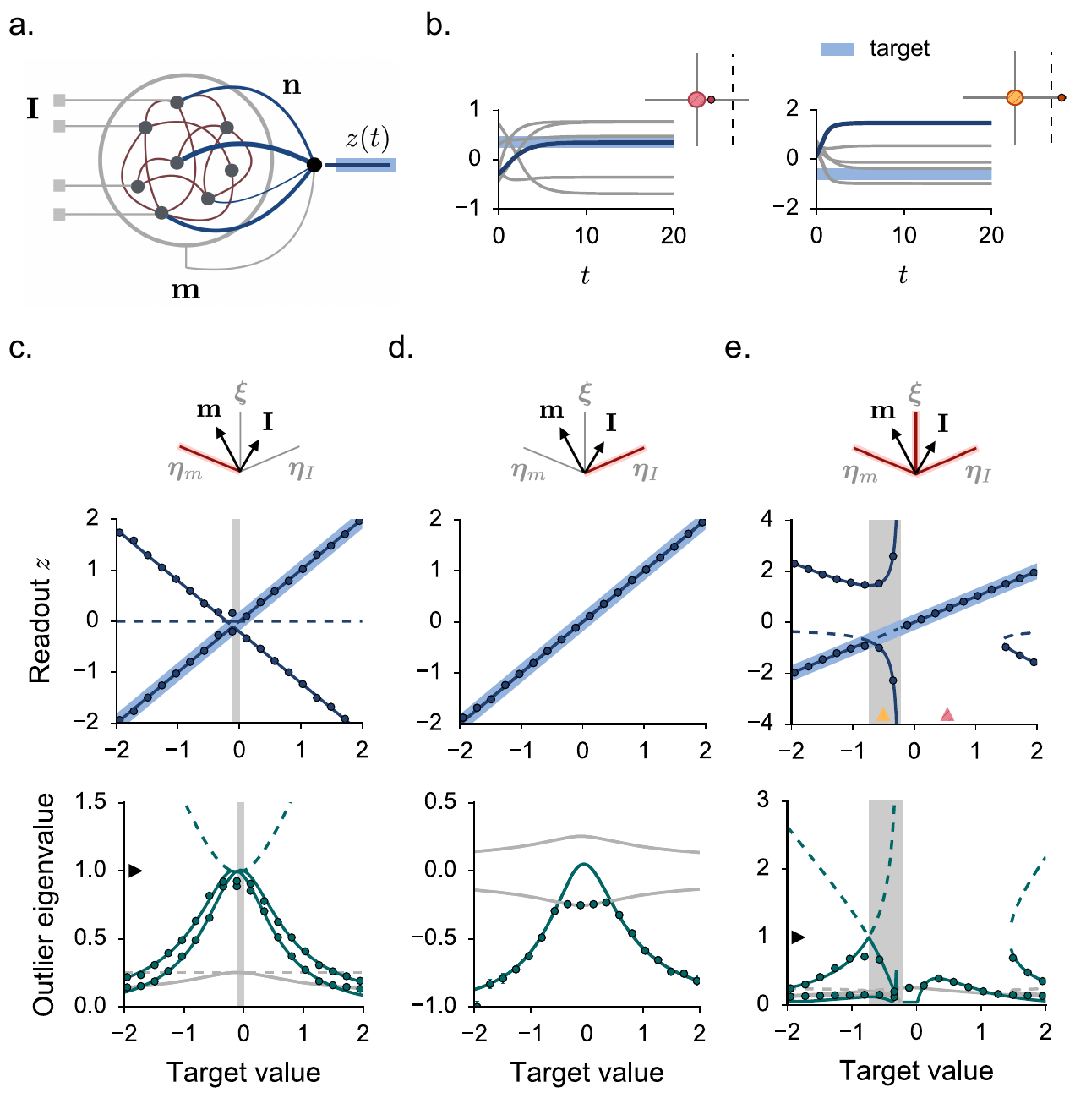}
	
	\caption{ Implementing the fixed point task: network model and mean-field analysis in three example geometries.
		{\bfseries a.} Architecture of the feedback network (Eq.~\ref{eq:full_dynamics_rankone}).
		{\bfseries b.} Sample from the dynamics of two finite-size networks, $N=1000$. In the left example, the output state which corresponds to the target fixed point is locally stable, and the readout signal $z(t)$ converges to the correct value. In the right example, instead, the target corresponds to a locally unstable state, and the readout converges to an incorrect value. Gray traces indicate the time course of the activation variable $\mathbf{x}$ for five randomly selected units. The readout trajectory is instead displayed in blue. 
		The top-right insets display the mean-field prediction for the stability eigenspectrum of the target fixed point (see \emph{Appendix A}). Local instabilities are due to the outlier eigenvalue crossing the instability boundary (dashed). The parameters which have been used in the left and the right examples are indicated by colored arrows in {\bfseries e}.
		{\bfseries c-d-e.} Mean-field characterization of the network fixed points for three different readout geometries (see text). Continuous (resp. dashed) lines correspond to locally stable (resp. unstable) mean-field solutions. Shaded areas indicate the values of the target $A$ for which the association is locally unstable. Top row: value of the readout signal. Bottom row: position of the outlier eigenvalue in the stability eigenspectra. Gray lines indicate the value of the radius of the eigenvalues bulk. The instability boundary is indicated on the ordinate axis by the black arrows.
		The results of simulations are displayed as dots ($N=3000$, average over 8 network realizations). We integrate numerically the dynamics of finite-size networks where the readout vector $\mathbf{n}$ is normalized through Eq.~\ref{eq:c}.
		In order to reach different solutions, we initialize the network dynamics in two different initial conditions, centered around $\mathbf{n}$ and $-\mathbf{n}$. Note that the position of the outlier eigenvalue cannot be measured when the outlier is absorbed in the circular set of the eigenspectrum. In order to ease the comparison between theory and simulations, we thus chose a small $g$ value.
		Parameters: $g=0.3$, $\sigma_m=1.2$, $\sigma_I=0.5$. $\rho=0.5$. In {\bfseries e}, we take $p=p_m = 1$, $p_I=0.3$. Here and in the following figures: error bars, when visible, correspond to the standard error.
	}
	\label{fig:general}
	
\end{figure}

Once the normalizing factor has been derived, the full mean-field equations can be solved. Because of the normalization, the mean-field system of equations always admits a stationary solution which satisfies the task for $N \rightarrow \infty$ (i.e.~its readout obeys $z=A$). This solution, however, is not guaranteed to be locally stable from the point of view of the full feedback dynamics (Fig.~\ref{fig:general} {\bf b}). Apart from the target solution, furthermore, the mean-field equations might admit other solutions corresponding to stable fixed points.
In those cases, depending on the initial conditions, the dynamics of the final network might converge to spurious output states where the readout $z$ significantly deviates from the target.

For every solution of the mean-field equations, we predict local stability by evaluating the eigenspectrum of the linear stability matrix of the corresponding fixed point. For the class of readouts that we consider, the eigenspectra consist of a dense set of eigenvalues, distributed within a circle in the complex plane, together with a single real outlier (Fig.~\ref{fig:general} {\bf b}) \citep{MastrogiuseppeOstojic2, Tao2013, RivkindBarak}. The radius of the circular set and the position of the outlier eigenvalue can be evaluated within the mean-field framework \citep{MastrogiuseppeOstojic2}; their value is determined by the interplay between the amplitude of the random bulk and the relative geometrical arrangement of vectors  $\mathbf{n}$, $\mathbf{m}$ and $\mathbf{I}$.
Details of the calculations are provided in \emph{Appendix A}.

\subsection{Categories of readout solutions}

We specifically look at the results obtained for three readout vectors characterized by different geometries. 

In the first example we analyze (Fig.~\ref{fig:general} {\bf c}), the readout $\mathbf{n}$ overlaps solely with the feedback input $\mathbf{m}$ along $\bm{\eta}_m$. This configuration corresponds to setting $p = p_I = 0$, and can be realized also in absence of the external input vector. In this case, the network admits three stationary states for every value of the target $A$. Among them, only two states are locally stable. Bistability is due to the strong overlap between the feedback vectors $\mathbf{m}$ and $\mathbf{n}$, and emerges through a pitchfork bifurcation induced by the outlier eigenvalue in the stability eigenspectrum of the feedback-free fixed point \citep{Hopfield, MastrogiuseppeOstojic2}.
As a consequence, stationary states tuned to small target values are characterized by long relaxation timescales. 
For most of the target values, the solution corresponding to the target (i.e.~$z=A$) is locally stable. However, the network admits a second stable fixed point, characterized by a different readout value. This additional stable state is reached when the network is initialized within the appropriate basin of attraction. The position of the unstable fixed point (dashed lines in Fig.~\ref{fig:general}) can be used to estimate the size of the basins of attraction of the two stable fixed points, which in this case is approximately equal.
Within a small parameter region corresponding to very small target values, the target fixed point is built in correspondence of the intermediate, locally unstable branch of the mean-field solutions. The mean-field theory indicates that the local instability is due to the outlier eigenvalue of the fixed point stability matrix laying above the critical line (Fig.~\ref{fig:general} {\bf b} right).
The amplitude of the instability region is controlled by the overlap between the two fixed vectors $\mathbf{m}$ and $\mathbf{I}$, and disappears when $\rho = 0$ \citep{RivkindBarak}. 

In the second example we consider (Fig.~\ref{fig:general} {\bf d}),  the readout $\mathbf{n}$ overlaps solely with the external input weights $\mathbf{I}$ along $\bm{\eta}_I$. We thus set: $p = p_m = 0$. Analogously to a feedback-free network, aligning the readout vector with the external input generates a unique stationary solution, corresponding to the target of the task ($z=A$). Such a solution is always stable, and it is characterized by negative values of the outlier eigenvalue.

As a last example, we consider the general case where the readout displays non-vanishing overlaps with both vectors $\mathbf{m}$ and $\mathbf{I}$ (Fig.~\ref{fig:general} {\bf e}). Such a category includes the simplest configuration where the readout $\mathbf{n}$ is aligned with the overlap direction $\bm{\xi}$ ($p_m=p_I=0$), but more general solutions characterized by non-zero projections along $\bm{\eta}_m$ and $\bm{\eta}_I$ are comprised as well. 
As in the first example we considered (Fig.~\ref{fig:general} {\bf c}), the components of the readout which overlap with the feedback input $\mathbf{m}$ tend to generate three different fixed points. 
Because of the components along the external input $\mathbf{I}$, however, these solutions are no longer characterized by a  strong symmetry around zero.
As a result, the unstable intermediate fixed point is always characterized by finite output values. This unstable state corresponds to the task solution within a finite parameter region at small and negative target values. This instability area is delimited by a critical target value $A^*$ where the readout amplitude $c$, together with the stable fixed points, diverges:
\begin{equation}
A^* = - \frac{ p \sigma_I  \rho + p_I \sigma_I \sqrt{1-\rho^2}}{p \sigma_m \rho + p_m\sigma_m\sqrt{1-\rho^2}}.
\end{equation}
Note that this geometry produces a unique and stable input-output association within a parameter window corresponding to small and positive target values. For large (positive or negative) target values, instead, the projection along $\mathbf{m}$ dominates and a second stable fixed point exists.

To conclude, we characterized the output states generated by readout solutions in the class defined by Eq.~\ref{eq:sol}.  We found that, for different solutions belonging to that class, the task is satisfied through output states which are characterized by different local and global stability properties. Stability can be predicted via the mean-field theory by examining the projections of the readout $\mathbf{n}$ on the input vectors $\mathbf{m}$ and $\mathbf{I}$. In particular, an optimal solution -- characterized by both local and global stability -- can be obtained by aligning the readout $\mathbf{n}$ along the non-shared direction of the external input $\mathbf{I}$. This solution requires a non-vanishing input pattern and generates a feedback network that is dynamically analogous to the feedback-free implementation.
Components along the feedback input $\mathbf{m}$, on the other hand, tend to generate additional stable solutions which can attract the dynamics to different readout values. 
Combining both readout components along  $\mathbf{m}$ and $\mathbf{I}$, finally, generates non-symmetric bistable solutions which result in broad regions of local and global instability.

\section{\textsc{Least-Squares training}}

In Section 3, we derived a theoretical framework which allows us to directly map the geometry of the readout vector into the final output states of the feedback network. This description applies to networks which obey three closely related assumptions (Eq.~\ref{eq:sol}): (i) the readout is a linear combination of the input axes $\bm{\xi}$, $\bm{\eta}_{{m}}$, $\bm{\eta}_{{I}}$; (ii) no correlations exist  between the readout and the random bulk; (iii) the scaling of the solution with the network size $N$ is weak (i.e. $n_i$ is of order $1/N$).

Importantly, these assumptions might not hold in finite-size networks obtained through standard training techniques. Trained networks might indeed converge to more complex, non-linear solutions, where strong correlations with the random connectivity significantly influence the output dynamics.
In Sections 4 and 5, we investigate how far our approximate theory can be used to describe trained networks. 
To this end, we construct for every training solution a crude approximation in the form of Eq.~\ref{eq:sol}, for which the mean-field equations can be solved and the global dynamics of the resulting network can be predicted. 
Since in the mean-field description correlations with the random bulk are neglected, comparing mean-field predictions with training performance allows us to quantify how much training relies on a fine-tuning of the readout to the random connectivity $\bm{\chi}$. In cases where the theory predicts that no stable solutions exist, furthermore, our analysis allows to investigate whether finite-size trained networks converge to more effective solutions that the simplified theoretical description cannot capture.

We consider in two specific cases, corresponding to two classical training protocols.
The first training procedure we consider is the simple batch update through least-squares (LS) inversion \citep{Jaeger, Lukosevicius, RivkindBarak}. 
To begin with, a finite-size architecture defined by a random bulk $\bm{\chi}$ and two vectors $\mathbf{m}$ and $\mathbf{I}$ is generated.
The activity of the whole population in the open-loop configuration is then simulated, yielding:
\begin{equation}\label{eq:xi_star}
x_i^{ol} = m_i A + I_i + \delta_i^{ol}.
\end{equation}
The task imposes a unique constraint on the readout, namely: $\phi(\mathbf{x}^{ol})^T \mathbf{n} = A$. This equation represents a underdetermined linear system for the $N$ entries of vector $\mathbf{n}$. 
The minimum norm LS solution can be computed through the pseudo-inverse of $\phi(\mathbf{x}^{ol})^T$ \citep{Bretscher}:
\begin{equation}\label{eq:LS}
\mathbf{n} = \frac{A}{N \langle \phi^2(x_i^{ol})\rangle}\phi(\mathbf{x}^{ol})
\end{equation}
and thus scales with the network size as $1/N$.

Through the open-loop activation vector $\mathbf{x}^{ol}$,  the LS readout inherits three $N$-dimensional components along vectors $\mathbf{m}$, $\mathbf{I}$ and $\bm{\delta}^{ol}$ (Eq.~\ref{eq:xi_star}). The non-linear activation function $\phi(x)$, which is applied to $\mathbf{x}^{ol}$ (Eq.~\ref{eq:LS}), mixes and reshapes these three directions, by generating a continuum of additional components along orthogonal axes.
The component of $\mathbf{x}^{ol}$ aligned with the recurrent input $\bm{\delta}^{ol}$, furthermore, introduces finite, $\mathcal{O}(1)$ correlations between the readout and the finite-size instantiation of the random bulk used for training. Note that a fraction of those correlations is used by the LS algorithm to exactly match the readout $z$ to the target $A$ within network instantiations of any size. Mean-field solutions (Eq.~\ref{eq:sol}), instead, are independent of the random bulk and imply small readout errors (of the order of $1/\sqrt{N}$) which disappear only in the thermodynamic limit.

\subsection{Mean-field approximation}

We construct an approximate description of $\mathbf{n}$ (Eq.~\ref{eq:LS}) which conserves the relative strength of its components along the input vectors $\mathbf{m}$ and $\mathbf{I}$ (Eq.~\ref{eq:sol}), thus totally neglecting the components originating from the recurrent input $\bm{\delta^{ol}}$ and the non-linearity. 
To this end, we extract the linear projections of the readout vector along the orthogonal axes $\bm{\xi}$, $\bm{\eta}_m$ and $\bm{\eta}_I$, resulting in a set of weights $(p, p_m, p_I)$ which effectively replaces the LS solution with an approximate readout in the form of Eq.~\ref{eq:sol}. For such an approximate readout, the network output states can be exactly predicted.

As we show in the next paragraphs, corrections to the mean-field description due to both neglected terms $\bm{\delta^{ol}}$ and $\phi(x)$ can, in this case, be exactly computed \citep{RivkindBarak}.
A minimal approximation is however of interest per-se, as it directly generalizes to more complex readout vectors for which analytical expressions do not exist, so that the exact form of correlations and the shape of non-linear components is not known a priori (see Section 5).

The projection of $\mathbf{n}$ on $\bm{\xi}$ can be computed as:
\begin{equation}\label{eq:LS_alpha}
\begin{split}
p &=\frac{A}{N \langle \phi^2(x_i^{ol})\rangle} \langle \xi_i \: \phi(x_i^{ol}) \rangle\\
& =\frac{A}{N \langle \phi^2(x_i^{ol})\rangle} \langle \xi_i \: \phi\left( \sigma_m(\rho \xi_i + \sqrt{1-\rho^2}\eta_{mi}) A +\sigma_I(\rho \xi_i + \sqrt{1-\rho^2}\eta_{Ii}) + \delta_i^{ol}\right)\rangle.
\end{split}
\end{equation}
We re-express the population average as a Gaussian integral and we integrate by parts over $\xi_i$, yielding:
\begin{equation}
p = \frac{A }{N \langle \phi^2(x_i^{ol})\rangle}\rho(\sigma_mA + \sigma_I ) \langle \phi'(x_i^{ol}) \rangle .
\end{equation}

Similar calculations return the values of $p_m$ and $p_I$. The final set of weights is given by:
\begin{equation}\label{eq:LS_weights}
\begin{split}
& p = \gamma \rho(\sigma_mA + \sigma_I ) \\
& p_m = \gamma \sigma_m  \sqrt{1-\rho^2}A \\
& p_I  = \gamma \sigma_I  \sqrt{1-\rho^2} ,
\end{split}
\end{equation}
where we defined the common multiplicative factor $\gamma = A \langle \phi'(x_i^{ol}) \rangle / N \langle \phi^2(x_i^{ol})\rangle $.

As in Fig.~\ref{fig:general} {\bf e}, the LS solution thus includes non-vanishing components along the three axes $\bm{\xi}$, $\bm{\eta}_m$ and $\bm{\eta}_I$. In contrast to Fig.~\ref{fig:general}, however, the values of $(p, p_m, p_I)$ that have been derived from the LS solution display an explicit dependence on the target value $A$. For fixed vectors $\mathbf{m}$ and $\mathbf{I}$, the geometrical arrangement of the readout vector thus changes according to the target of the task.

\begin{figure}[t]
	\centering
	
	\includegraphics{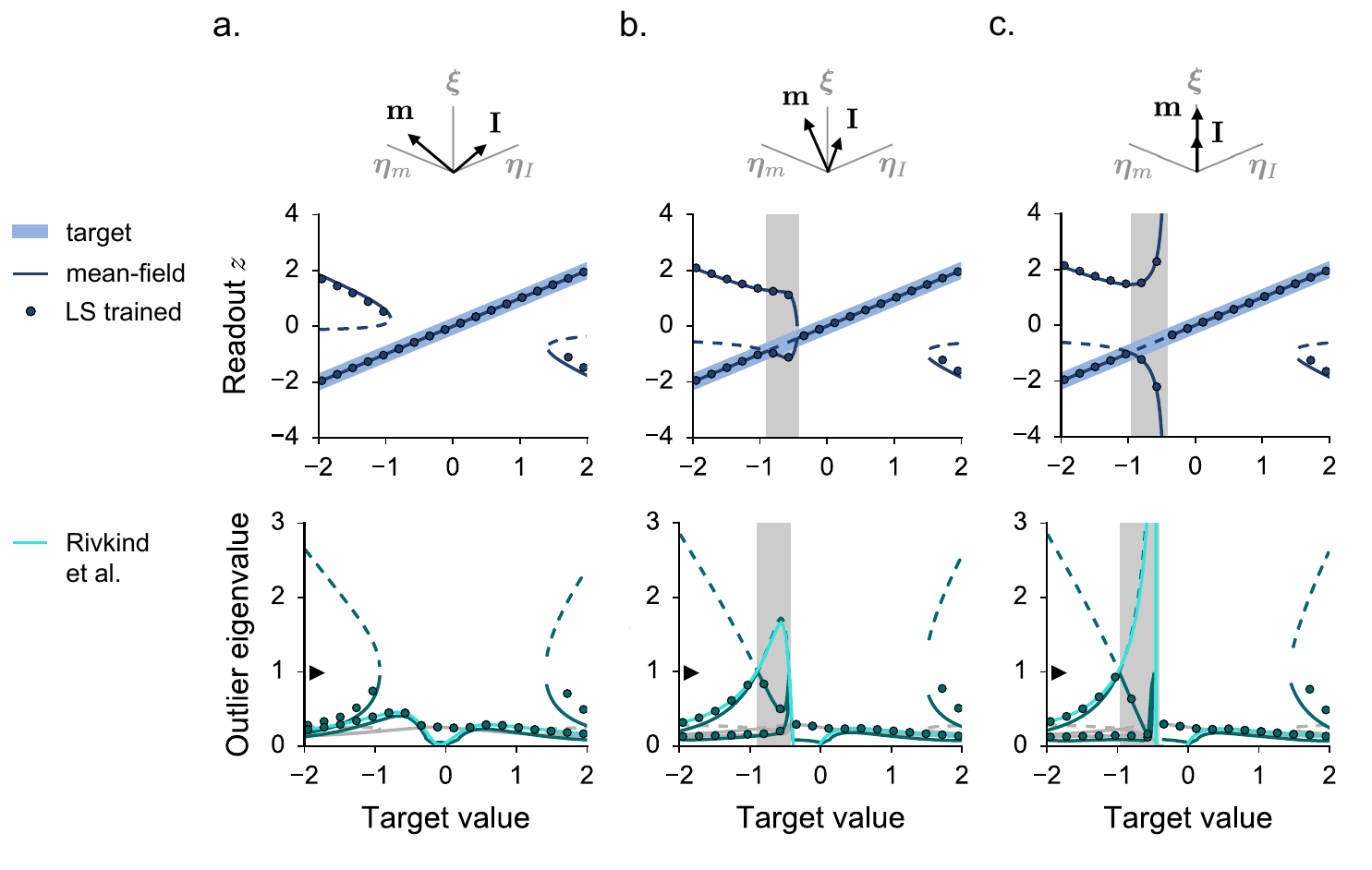}
	
	\caption{ 
		The least-squares solution: approximate mean-field description. 
		{\bf a-b-c.} Mean-field characterization for three different configurations of the non-trained input vectors $\mathbf{m}$ and $\mathbf{I}$. In {\bf a}, we take $\rho=0.6$; in {\bf b}, $\rho=0.97$; in {\bf c}, $\rho=1$. Continuous and dashed lines indicate the solutions of the simplified mean-field description, where the readout is approximated by a vector in the form of Eq.~\ref{eq:sol}. Details are as in Fig.~\ref{fig:general}. The light green line indicates the exact value of the outlier eigenvalue which measures the local stability of the target solution \citep{RivkindBarak}; details are provided in \emph{Appendix B}. 
		Dots display the results of simulations from finite-size networks ($N=3000$, averages over 8 network realizations). We integrate numerically the dynamics of finite-size networks where the readout vector $\mathbf{n}$ is given by the LS solution (Eq.~\ref{eq:LS}). In order to reach different solutions, we initialize the network dynamics in two different initial conditions, centered around $\mathbf{n}$ and $-\mathbf{n}$. 
		Parameters as in Fig.~\ref{fig:general}. 
	}
	\label{fig:LS}
	
\end{figure}

Starting from Eq.~\ref{eq:LS_weights}, we evaluate the normalization factor (Eq.~\ref{eq:c}) and we solve the system of mean-field equations
(Eqs.~\ref{eq:mf}, \ref{eq:kappa_fin}). In Fig.~\ref{fig:LS}, the solutions and their stability properties are illustrated.

We find that the shape of the mean-field solutions critically depends on the geometrical arrangement of the non-trained part of the network architecture.
Specifically, the network output is strongly influenced by the value of the overlap between the input vectors $\mathbf{m}$ and $\mathbf{I}$, quantified by the parameter $\rho$.

When the input vectors $\mathbf{m}$ and $\mathbf{I}$ are orthogonal or share a weak overlap, the theory admits a continuous stable solution -- on which the target output is built -- together with two symmetric stable branches at large target values (Fig.~\ref{fig:LS} {\bf a}). In this case, the theory predicts that training always generates locally stable dynamics \citep{RivkindBarak}, but spurious fixed point attractors are generated at large $A$.
As the overlap $\rho$ is increased, the branch of the solution corresponding to the target merges with one of the two additional fixed points, generating a region of local instability at small and negative target values (Fig.~\ref{fig:LS} {\bf b}).
In the limit where the input vectors $\mathbf{m}$ and $\mathbf{I}$ are fully aligned ($\rho=1$, Fig.~\ref{fig:LS} {\bf c}), finally, mean-field solutions are similar to the results in Fig.~\ref{fig:general} {\bf e}. The network is characterized by a region of local instability which is delimited by a critical target value $A^*$ where the dynamics diverge. The normalization factor $c$ is in this case given by:
\begin{equation}
c = \frac{A}{ \rho \left\{\sigma_I +  \sigma_m A \right\}^2 \langle [\phi'(x_i^{ol})]\rangle} ,
\end{equation}
which diverges at $A^* = - \sigma_I/\sigma_m$.

In Fig.~\ref{fig:LS}, mean-field predictions are compared with the outcome of simulations performed in finite-size networks which have been trained through LS inversion. Despite the approximate nature of the theory, a good agreement between the two is found.
As predicted by the theory, output states of trained networks are strongly affected by the value of the input overlap $\rho$. In particular, the theoretical approximation correctly predicts the existence of an area of local instability for large  $\rho$ values. Within this parameter region, training generates an output $z(t)$ which immediately diverges from the desired target.
For every value of $\rho$, furthermore, the theory captures the existence of a second, spurious fixed point at large target values. 

The mean-field prediction for the outlier stability eigenvalue can be further compared with an exact analytical expression derived by following the analysis in \citet{RivkindBarak}. This expression, provided in \emph{Appendix B}, includes the components of the readout $\mathbf{n}$ which are neglected in the simplified mean-field description: correlations with the bulk $\bm{\chi}$ and non-linear geometry. We find that, apart from minor quantitative discrepancies, the exact outlier prediction is well matched by the approximate one. Note that this exact analytical expression only provides a measure of local stability, as it can be evaluated only in correspondence to the target solution.

To conclude, we found that dynamics in LS trained networks can be explained by a highly simplified, approximate theoretical description which conserves the projection of the readout vector on the hyperplane spanned by the feedback and external input vectors.
The approximate mean-field theory specifically predicts that LS readouts often generate bistable fixed points, while unstable solutions can be encountered for large values of the overlap $\rho$. Note that, as long as the input vectors $\mathbf{m}$ and $\mathbf{I}$ are not completely parallel (i.e.~$\rho<1$), the mean-field framework indicates that an optimal readout solution can be designed by aligning the readout $\mathbf{n}$ with the non-shared component of $\mathbf{I}$ (Fig.~\ref{fig:general} {\bf d}).
We conclude that, even for the simple task we consider, LS training does not lead to an optimal solution that prevents dynamical instabilities. 

Apart from an overall quantitative agreement, mismatches between simulations and mean-field solutions can be observed. Their size depends on the value of the target and on the network parameters. A brief description of the role and impact of the different sources of mismatch is presented in the next two paragraphs.

\subsection{Non-linear corrections}

\begin{figure}
	
		\centering
		\includegraphics{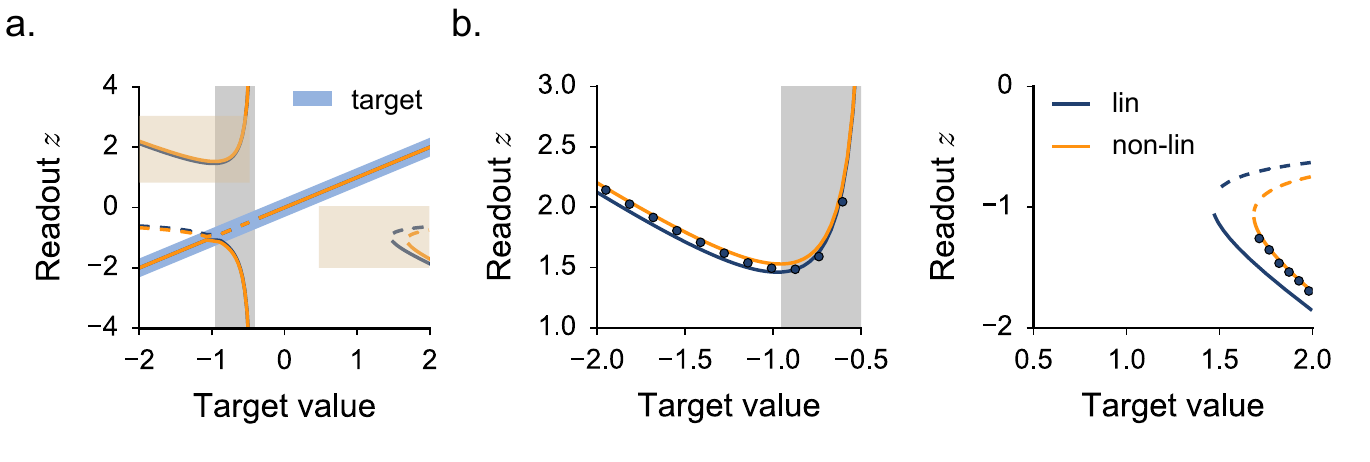}
	
	\caption{ 
		The least-squares solution: mean-field description with non-linear corrections ($\rho=1$). 
		{\bf a.} We compare the solutions which are obtained from the linear (blue) and the non-linear (orange) set of mean-field equations. Note that the two set of solutions almost completely overlap. 
		{\bf b.} We plot a magnified view of the solutions within the shaded yellow areas of {\bf a}. We compare both mean-field solutions with the results of simulations performed in LS-trained networks (dots). Parameters and simulations settings are as in Fig.~\ref{fig:LS}.
	}
	\label{fig:nonlinear}
	
\end{figure}

Together with its components along the input axes $\bm{\xi}$, $\bm{\eta}_m$ and $\bm{\eta}_I$, the full LS readout (Eq.~\ref{eq:LS}) includes a continuum of orthogonal directions which can potentially affect the network dynamics and output values.
We first focus on the orthogonal components which are generated when the non-linear function $\phi(x)$ is applied to $\mathbf{x}^{ol}$  (Eq.~\ref{eq:LS}).

We build a more precise mean-field approximation which conserves the non-linearity of the LS solution (Fig.~\ref{fig:nonlinear}). Details of the analysis are provided in \emph{Appendix D}.
By solving the novel set of mean-field equations, we altogether find that non-linear solutions lie very close to the ones we obtained for linearly approximated readouts (Fig.~\ref{fig:nonlinear} left). The simple linear approximation, therefore, appears to capture well the readout geometry that is relevant for the final network dynamics. At a finer scale, we find that non-linear solutions in some cases explain with higher precision the output states observed in trained networks. In particular, we observe that the agreement significantly improves in the parameter regions corresponding to large and positive target values, where the dynamics are strongly non-linear and admit bistable fixed points (Fig.~\ref{fig:nonlinear} right).

\subsection{The effect of correlations}

Our approximate mean-field description neglects a second set of orthogonal readout directions, generated by the recurrent input $\bm{\delta}^{ol}$ (Eq.~\ref{eq:xi_star}). When this vector is replaced with Gaussian noise, correlations between the LS readout and the random part of the connectivity are effectively washed out.

\begin{figure}
	
	\begin{adjustwidth}{-0.3in}{0in} 
		\centering
		\includegraphics{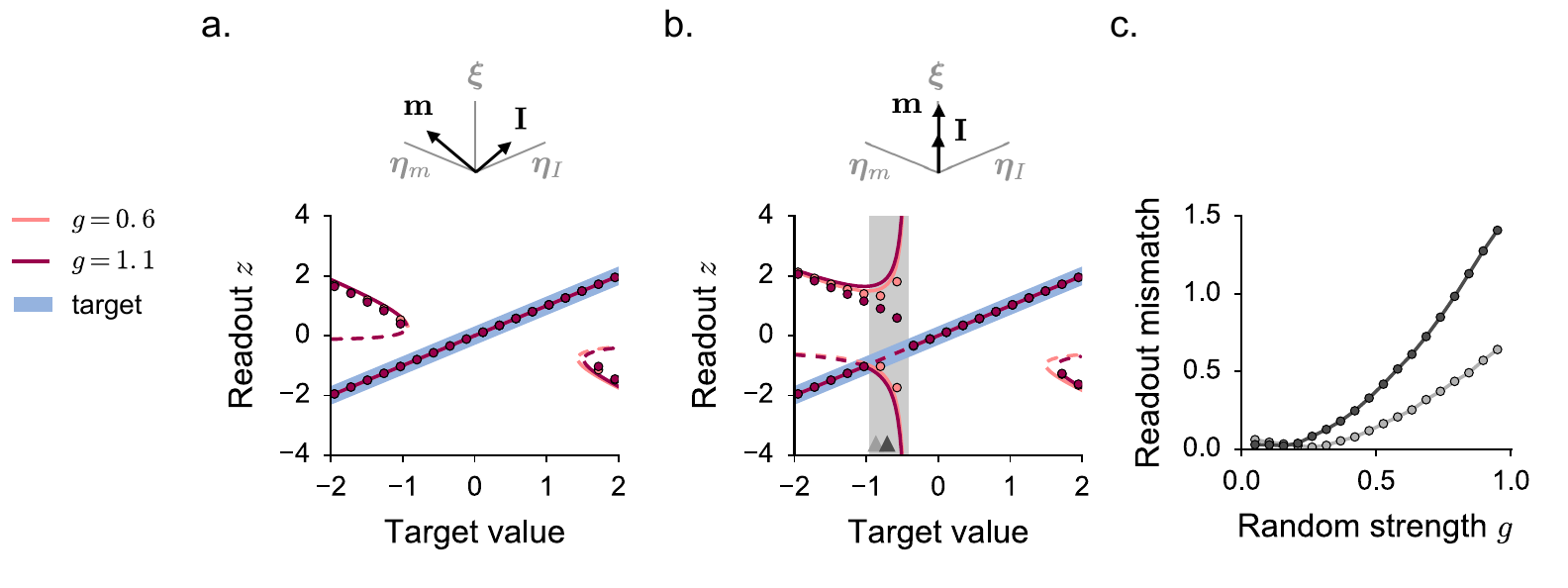}
	\end{adjustwidth}
	
	\caption{ 
		The least-squares solution: effect of correlations.
		{\bf a-b.} Comparison between the linear mean-field predictions and the trained networks output states for two increasing values of $g$, corresponding to pink ($g=0.6$) and magenta ($g=1.1$) traces. 
		We consider network architectures characterized by two different values of the input vectors overlap:
		in {\bf a}, we take $\rho = 0.6$, in {\bf b}, $\rho=1$.  Note that, especially in {\bf a}, the results for the two values of $g$ strongly overlap.
		A significative mismatch between theory and simulations is observed in {\bf b}, for $g=1.1$, within the instability area at small and negative target values.
		{\bf c.} For $\rho=1$, we fix the value of the target $A$ within the instability window (grey arrows in {\bf b}) and we measure the mismatch between mean-field predictions and trained networks as the random strength $g$ is increased. The mismatch is measured at the level of the readout $z$ in correspondence of the spurious fixed point characterized by positive readout values.
		Parameters and simulations settings are as in Fig.~\ref{fig:LS}.
	}
	\label{fig:correlations_a}
	
\end{figure}

In order to boost the effect of correlations, we increase the strength of random connections $g$ up to the critical point where dynamics become chaotic (Fig.~\ref{fig:correlations_a} {\bf a-b}; note that the rank-one structure and the external input pattern shift the critical coupling from $g=1$ to slightly larger values \citep{Rajan2010, Schuecker2017}). For every value of $g$, we observe that the approximate mean-field description correctly captures performance in LS-trained networks when the two input vectors $\mathbf{m}$ and $\mathbf{I}$ are not completely aligned ($\rho<1$, Fig.~\ref{fig:correlations_a} {\bf a}). On the other hand, we find that the theory fails to fully describe degenerate architectures with parallel input vectors ($\rho=1$, Fig.~\ref{fig:correlations_a} {\bf b}): a significant qualitative mismatch is observed within the instability region corresponding to small and negative target values. In such a region, non-linear corrections have very little effect (Fig.~\ref{fig:nonlinear}), and the mismatch is mostly due to correlations between the readout vector and the random bulk. The mismatch indeed increases with the random strength $g$, as the recurrent input deriving from the random connectivity becomes large with respect to the feedback input (Fig.~\ref{fig:correlations_a} {\bf c}).

\begin{figure}[t]
	\begin{adjustwidth}{-0.3in}{0in} 
		\centering
		\includegraphics{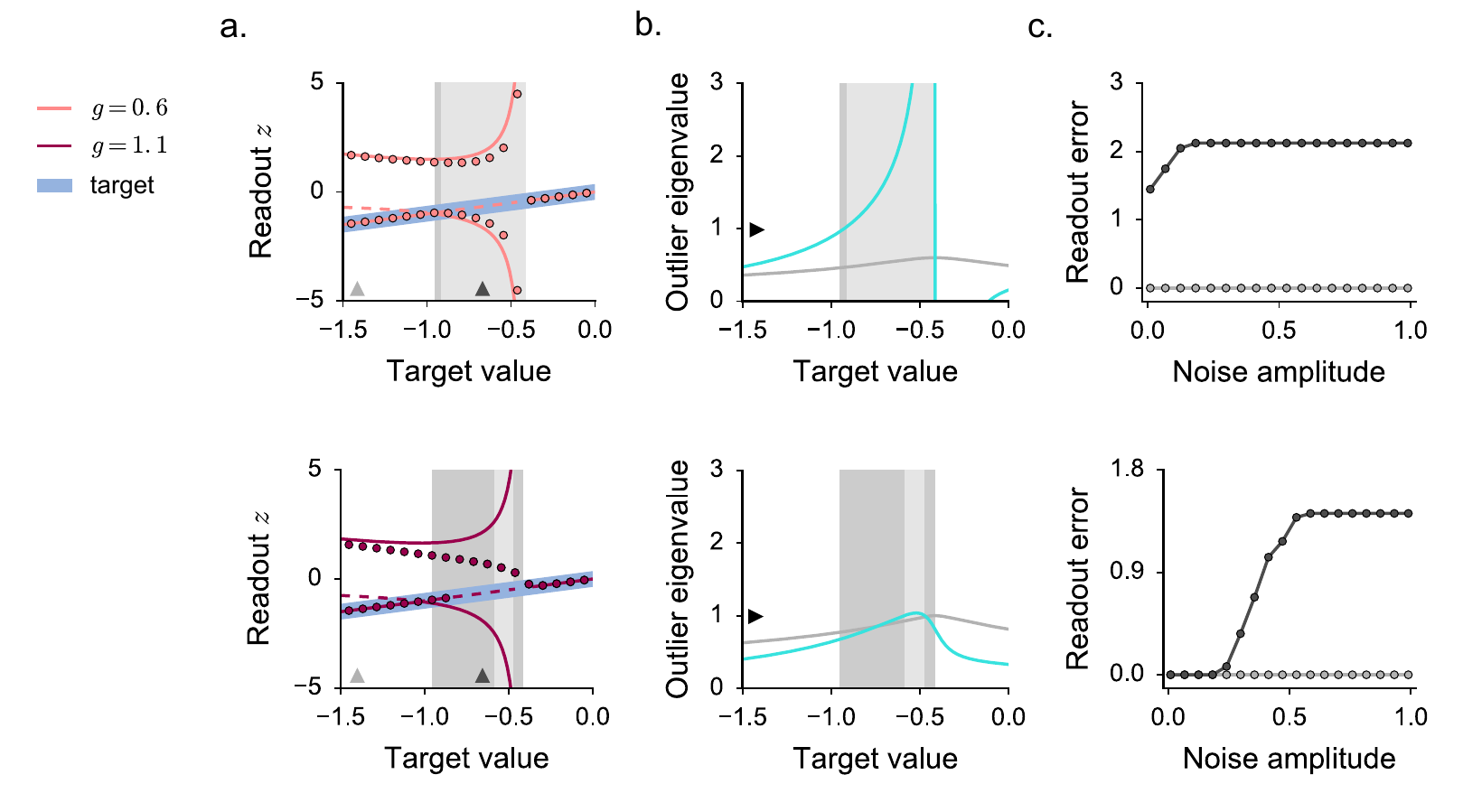}
	\end{adjustwidth}
	
	\caption{  
		The least-squares solution: effect of correlations; detailed analysis of the case $\rho=1$ (Fig.~\ref{fig:correlations_a}). The results for the two different values of $g$ are displayed separately in the top  ($g=0.6$) and in the bottom  ($g=1.1$) panels. 
		{\bf a.} Magnified view of Fig.~\ref{fig:correlations_a} {\bf b}. Arrows on the abscissa axis indicate the target values used in {\bf c}.  
		{\bf b.} Exact prediction for the outlier eigenvalue of the stability matrix of the target fixed point (in green, see \emph{Appendix B}). The outlier eigenvalue lies above the instability boundary within a parameter region corresponding to exact instability area, indicated in light gray. The instability area predicted by the approximate mean-field theory is instead indicated in dark gray.
		Gray lines indicate the value of the radius of the eigenvalues bulk.
		{\bf c.} Probing the basin of attraction of the target solution for two different values of $A$, indicated by arrows in {\bf a}. We simulate the post-training network dynamics starting from initial conditions given by Gaussian vectors of variable amplitude added on top of the open-loop activity $\mathbf{x}^{ol}$. The absolute error is then measured at the level of the readout $z$, and is averaged over 40 trained networks.
		Parameters are as in Fig.~\ref{fig:LS}.
	}
	\label{fig:correlations_b}
	
\end{figure}

A detailed analysis of degenerate feedback architectures ($\rho=1$) is provided in Fig.~\ref{fig:correlations_b}.
In the top row, we select an intermediate value of $g$. In this case, dynamics in trained networks diverge at the border of the instability window as predicted by the mean-field theory (Fig.~\ref{fig:correlations_b} {\bf a}).
In the bottom row, we consider a larger value of  $g$, laying right below the instability to chaos. In this case, the trained network output appears to obey a different qualitative pattern: the largest stable solution, which is predicted to diverge at $A^*$, converges instead at the border of the instability window.

Our approximate mean-field description predicts furthermore a large instability window, wider than the instability region which is predicted by the exact analytical expression \citep{RivkindBarak} (light shaded regions in Fig.~\ref{fig:correlations_b} {\bf a} and {\bf b}).
The target fixed point remains thus locally stable within a parameter region where the approximate theory predicts instability. Within the same window, however, the target solution is characterized by a stability eigenvalue laying very close to the instability boundary and a very narrow basin of attraction  (Fig.~\ref{fig:correlations_b} {\bf b} and {\bf c}).
Trained networks, indeed, converge to the target fixed point only when the initial conditions of the dynamics are very carefully tuned to the open-loop activation vector $\mathbf{x}^{ol}$. 
Since the dynamics systematically diverge from the target solution when the initial conditions are perturbed along random directions,  Fig.~\ref{fig:correlations_b} {\bf c} indicates that the basin of attraction of such a fixed point is particularly narrow with respect to any axis of the phase space. 

To conclude, we found that correlations impact the agreement between mean-field theory and trained networks within a restricted region of parameters, corresponding to almost parallel input vectors. In that region, a quantitative and qualitative match cannot be obtained when the ratio $g/A$, which measures the relative strength of the recurrent over the feedback dynamics, is large.
The approximate mean-field description, however, successfully captures the presence of an instability area, where the target fixed point is unstable or confined into a very narrow basin of attraction. 

\section{\textsc{Recursive Least-Squares training}}

Our analysis of least-squares solutions revealed the presence of a finite parameter region where the target fixed point is constructed on a locally unstable state. 
Dynamical feedback instabilities have been observed in a variety of studies where, as in the LS case we examined so far, the readout vector $\mathbf{n}$ is fixed through a unique and batch weights update \citep{Jaeger, Lukosevicius, Reinhart}.

A convenient strategy to overcome local instabilities consists of training feedback networks with an online algorithm so that the effect of synaptic changes is immediately propagated and tested on the network dynamics. FORCE (first-order reduced and controlled error) learning \citep{SussilloAbbott} is a popular example which relies on an online and recursive formulation of the least-squares algorithm (RLS) \citep{Liu, Jaeger2002}. This algorithm includes a regularization term, whose strength is controlled by a parameter $r$, which penalizes large amplitude readouts. Details of the training procedure are reported in \emph{Appendix E}.

In this section, we train feedback networks with FORCE and apply our theoretical framework to derive an approximate but effective description of the reached solutions.
Similarly to the previous section, we can investigate how the algorithm implements the task by replacing the trained readout vector with an approximation in the form of Eq.~\ref{eq:sol}. Specifically, we are interested in understanding whether -- and eventually, how -- RLS solutions overcome the local instabilities that are encountered by the batch version of the LS algorithm. 

\subsection{Parallel geometry}

We start our analysis from a specific case, corresponding to a particularly simple network architecture: the two input vectors $\mathbf{m}$ and $\mathbf{I}$ are taken to be parallel ($\rho=1$).
In this case, the mean-field theory for uncorrelated solutions (Eq.~\ref{eq:sol}) predicts that a unique solution exists, for which the readout vector $\mathbf{n}$ is parallel to the overlap axis $\bm{\xi}$, while orthogonal readout directions do not affect the output $z$. As in Figs.~\ref{fig:general} {\bf e} and \ref{fig:LS} {\bf c}, such a readout unavoidably generates regions of local and global instability.
Here we ask whether RLS solutions implement the task through a similar strategy, or whether alternative and more effective solutions are found outside of the restricted class of readouts which are described by the mean-field theory (Eq.~\ref{eq:sol}).

\begin{figure}[t]

		\centering
		\includegraphics{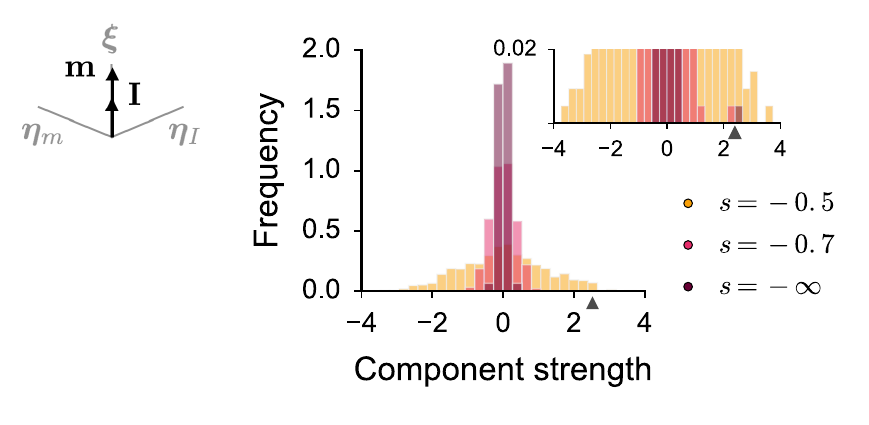}
	
	\caption{ 
		Recursive least-squares solutions ($\rho=1$): analysis of the readout vector.
		We consider a set of $N$ unitary and orthogonal vectors $\{\mathbf{v}^k\}_{k=1,..,N}$ which includes the input direction $\bm{\xi}$, and we decompose the trained readout $\mathbf{n}$ on such an orthonormal basis. We construct an histogram by collecting the strengths of the projections of $\mathbf{n}$ along the different basis vectors.
		Different histogram colors indicate three different training trials, where the norm of the initial guess for $\mathbf{n}$ has been varied.  The initial guess is generated as a Gaussian random vector of std $N^s$. The value of $s$ thus controls the amplitude of the initial guess with respect to the network size.  
		The inset reports a magnified view of the histogram for small ordinate values. The grey arrow indicates the average strength of the component of $\mathbf{n}$ along the input axis $\bm{\xi}$. The amplitude of the projections along the remaining basis vectors, instead, is strongly modulated by the initial amplitude $s$. 	Parameters: $g=0.3$, $\sigma_m=\sigma_I=1.2$, $\rho=1$, $A=1.6$. 
		}

	\label{fig:RLS_parallel}
	
\end{figure}

To start with, we run the RLS algorithm up to convergence and we then analyze numerically the geometry of the reached solution $\mathbf{n}$ (Fig.~\ref{fig:RLS_parallel}).
We find that, regardless of the network size, the value of the parameters and the initial conditions, the readout vector is systematically characterized by a large component aligned with the input direction $\bm{\xi}$ (grey arrow in the inset of Fig.~\ref{fig:RLS_parallel}). The readout vector, however, includes also strong components oriented in orthogonal directions: such additional components are induced by non-linearities, correlations with the random bulk and initial conditions of training.
In particular, since training convergence is often very fast, a large fraction of orthogonal components is directly inherited from the geometry of the initial guess for the readout $\mathbf{n}$, which is typically generated at random (Fig.~\ref{fig:RLS_parallel}). As a consequence, the amplitude of orthogonal components is strongly affected by the norm of the random vector which is used as initial guess for $\mathbf{n}$.
In Fig.~\ref{fig:RLS_parallel}, the initial readout is generated as a Gaussian vector of standard deviation $N^{s}$, so that the amplitude of orthogonal components is controlled by parameter $s$, which measures the amplitude of the initial guess with respect to the bulk size.

\begin{figure}[t]
	\begin{adjustwidth}{-0.3in}{0in} 
		\centering
		
		\includegraphics{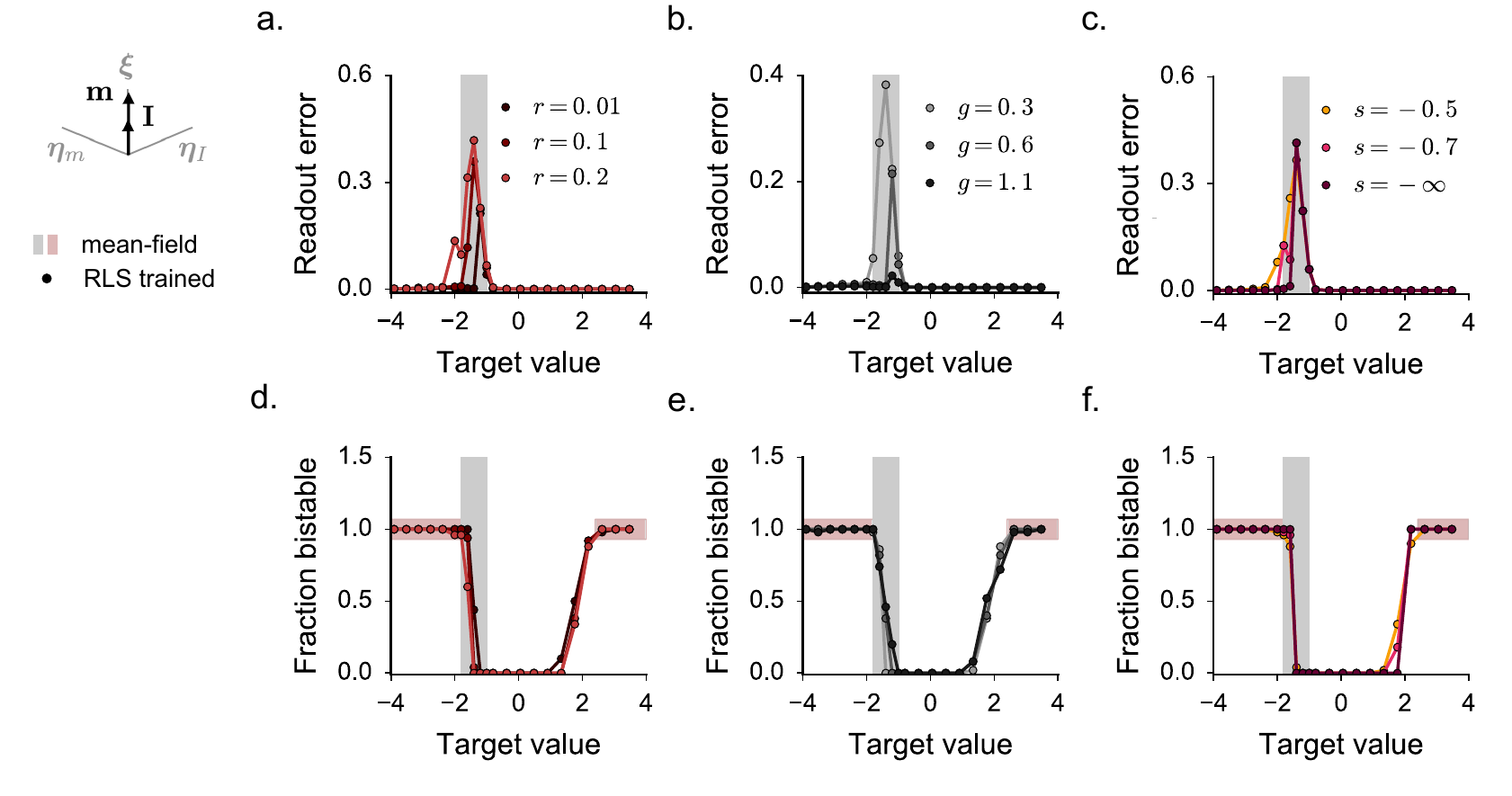}
	\end{adjustwidth}
	
	\caption{ 
		Recursive least-squares solutions ($\rho=1$): comparison between mean-field predictions and trained networks. The region of local instability predicted by the theory is displayed in gray. The regions where the mean-field theory predicts the existence of a second stable fixed point are instead indicated in {\bf d}, {\bf e} and {\bf f} by horizontal red stripes.
		In {\bf a}, {\bf b} and {\bf c}, we quantify the local stability of the training solution. To this end, the post-training activity is simulated for 50 normalized time units, where the initial condition is set by the activation variable during the last training step $\mathbf{x}^{end}$. The error is finally measured as $|z-A|$. 
		In {\bf d}, {\bf e} and {\bf f}, we test whether the network admits two bistable states. We simulate dynamics starting from $\mathbf{x}^{end}$ and $-\mathbf{x}^{end}$ as initial conditions ($\mathbf{n}$ and $-\mathbf{n}$ give similar results), and we compare the final value of the readout $z$ in the two cases. The trained dynamics are considered to be bistable if these two values have opposite sign.
		Dots show results averaged over 50 network realizations of size $N=600$. Here and in the following, continuous lines do not indicate analytical results, but are drawn to guide the eye.
		{\bf a, d.} Training results for different values of the regularization parameter $r$ (details provided in \emph{Appendix E}). 
		{\bf b, e.} Training results for different values of the random strength $g$. 
		{\bf c, f.} Training results for different values of the scaling $s$ of the initial guess for the readout $\mathbf{n}$. 
		Parameters as in Fig.~\ref{fig:RLS_parallel}.
	}
	\label{fig:RLS_parallel2}
	
\end{figure}

As a second step, we train feedback networks to different target values, and we attempt to predict performance through the help of the mean-field framework (Fig.~\ref{fig:RLS_parallel2}). As in the previous section, our approximate theoretical description conserves the component of $\mathbf{n}$ which is parallel to $\bm{\xi}$, and discards any orthogonal direction. In Eq.~\ref{eq:sol}, we thus set $p=1$, $p_m = p_I = 0$.

As in Figs.~\ref{fig:general} {\bf e} and \ref{fig:LS} {\bf c}, the mean-field theory predicts an area of local instability at small and negative target values (gray shaded regions in Fig.~\ref{fig:RLS_parallel2}). 
In order to evaluate the local stability of the fixed point generated by FORCE, we compute the post-training dynamics by using the activation vector from the last training step ($\mathbf{x}^{end}$) as an initial condition, and we finally measure the error at the level of the readout $z$.
Remarkably, we find that the RLS algorithm often fails to converge to a locally stable solution when the target $A$ is taken within the instability window predicted by the mean-field theory (Fig.~\ref{fig:RLS_parallel2} {\bf a}, {\bf b} and {\bf c}). The failure of the algorithm is thus correctly explained by our crude theoretical approximation, which replaces the complex readout emerging from training with a unique component along $\bm{\xi}$.

\begin{figure}[t]
	\centering
	\includegraphics{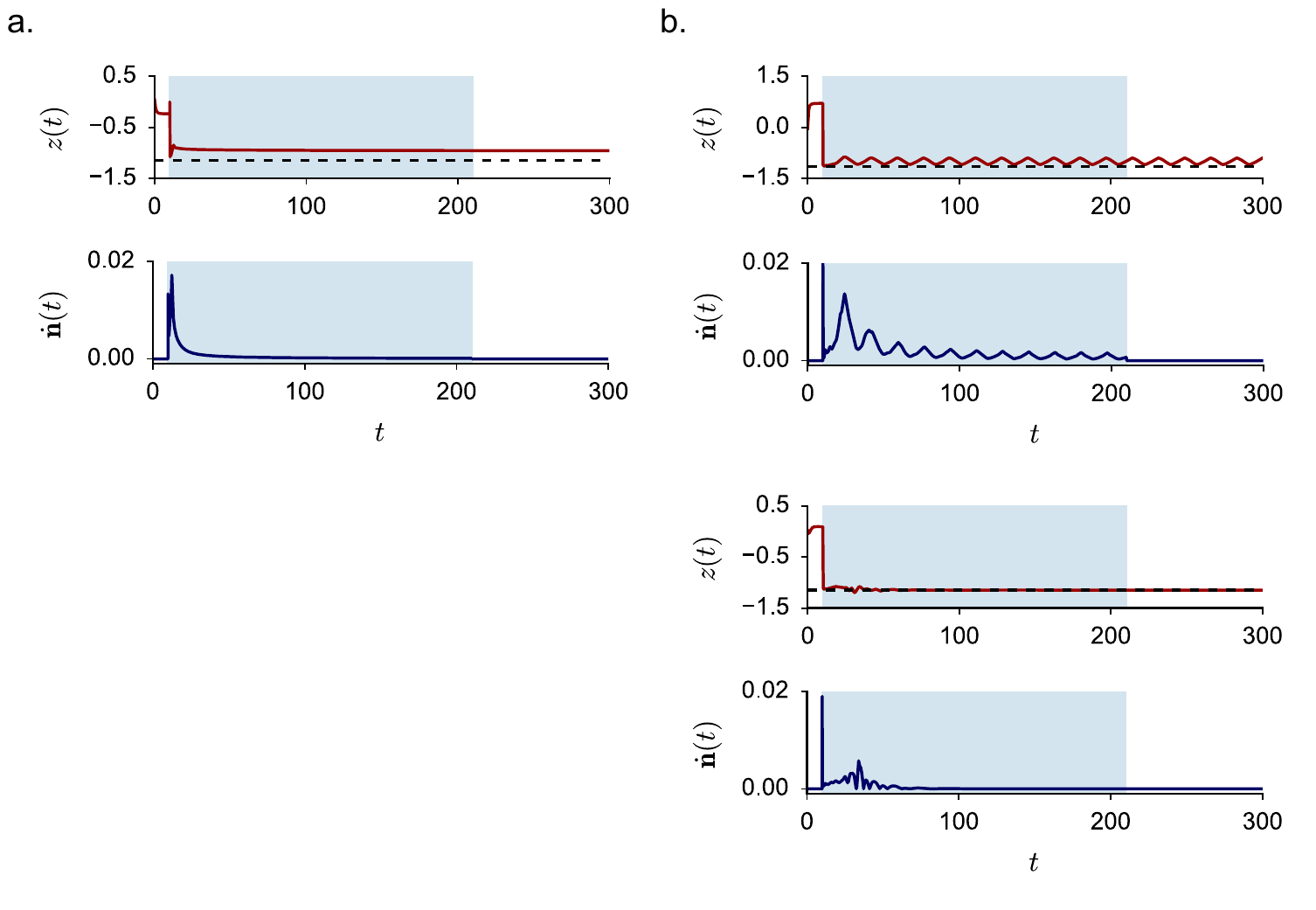}
	
	\caption{ 
		Recursive least-squares solutions ($\rho=1$): evolution of training for a target value within the instability window ($A=-1.1$). The top rows display the time course of the readout variable $z$, while the bottom rows show the amplitude of the synaptic modifications in $\mathbf{n}$ imposed by the algorithm.
		The training window corresponds to the shaded region, and the target $A$ is indicated by the dashed black line.
		In {\bf a}: $g=0.3$, in {\bf b}: $g=1.1$. In {\bf b}, we show a non-converging (top) and a converging (bottom) trial.
		Parameters as in Fig.~\ref{fig:RLS_parallel}. 
	}
	\label{fig:training}
\end{figure}

Although some variability exists, this result is robust with respect to a broad choice of network and training parameters. 
In particular, the average post-training error decays but remains positive for both small and large values of the random strength $g$ (Fig.~\ref{fig:RLS_parallel2} {\bf b}). 
Interestingly, training evolves differently in the two cases (Fig.~\ref{fig:training}). For small values of $g$, the algorithm never converges (Fig.~\ref{fig:training} {\bf a}). The readout error remains large during training, while the amplitude of the weights modification imposed by the algorithm smoothly decays. When the value of $g$ is large and close to the instability to chaos, instead, continuous weight modifications make the readout $z$ stably converge to the target within the majority of the trials (Fig.~\ref{fig:training} {\bf b} bottom), resulting in a small average error (Fig.~\ref{fig:RLS_parallel2} {\bf b}). These strongly correlated training solutions, which allow feedback networks to overcome the dynamical instability at small and negative target values, cannot be captured within our approximate theoretical framework.
We conclude that large $g$ values, corresponding to strong random connectivities, help the algorithm to find fine-tuned solutions whenever weak and uncorrelated readout vectors (Eq.~\ref{eq:sol}) cannot generate stable fixed points. 
In certain trials, however, the algorithm fails to converge (Fig.~\ref{fig:training} {\bf b} top), or generates fine-tuned solutions which diverge from the target after learning. As a consequence, the average post-training error never completely vanishes (Fig.~\ref{fig:RLS_parallel2} {\bf b}).

As in Figs.~\ref{fig:general} {\bf e} and \ref{fig:LS} {\bf c}, the mean-field approximation further predicts the existence of an additional stable fixed point for large -- positive and negative -- target values (red shaded regions in Fig.~\ref{fig:RLS_parallel2} {\bf d}, {\bf e} and {\bf f}). Such a fixed point is characterized by a readout value of opposite sign, and is responsible for large readout errors whenever the dynamics is initialized within its basin of attraction.
In order to assess bistability, we integrate the post-training dynamics starting from initial conditions centered around  $\mathbf{x}^{end}$ and $-\mathbf{x}^{end}$. Again, a good agreement with the mean-field prediction is found: trained networks converge to a second stable fixed point when the target value is taken within the parameter regions where the approximate mean-field network is bistable.

The systematic agreement between theory and simulations (Fig.~\ref{fig:RLS_parallel2}) altogether indicates that the readout component along $\bm{\xi}$ plays a major role in shaping the dynamical landscape of trained networks. Orthogonal readout directions, which are artificially included in the readout vector during training, can have strong amplitude (Fig.~\ref{fig:RLS_parallel}) but contribute little to the network output states.

\begin{figure}[t]
	
		\centering
		\includegraphics{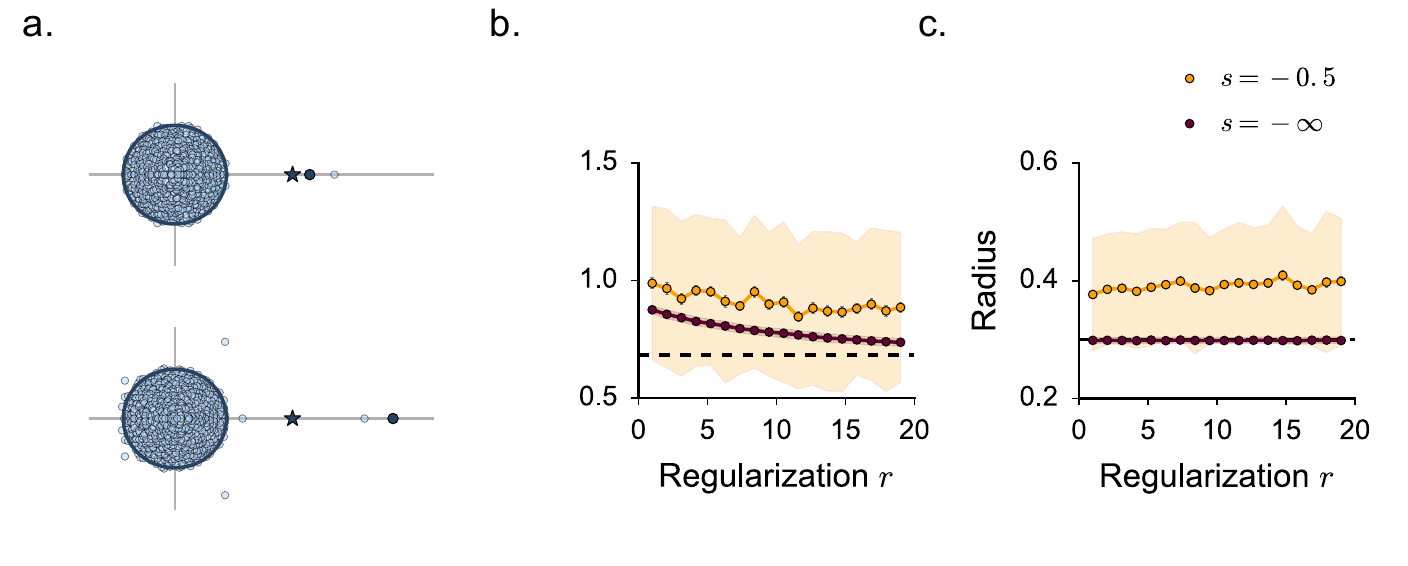}
	
	\caption{
		 Recursive least-squares solutions ($\rho=1$): analysis of the synaptic connectivity eigenspectra. 
		{\bf a.} Sample eigenspectra of $\bm{\chi} + \mathbf{mn}^T$. Top (resp. bottom) example: the initial guess for $\mathbf{n}$ is weak- (resp. strong-) amplitude: $s=-\infty$ (resp. $s=-0.5$). The dark blue dot indicates the position of the eigenvalue of the rank-one matrix $\mathbf{mn}^T$. The dark blue star indicates the position of the theoretical mean-field prediction. For a readout in the form of Eq.~\ref{eq:sol}, the two values coincide.
		{\bf b.} Simple estimate for the position of the outlier eigenvalue, measured as the largest real part across all the eigenvalues of the trained connectivity matrix.
		{\bf c.} Simple estimate for the radius of the circular set of eigenvalues, measured as the largest imaginary part across all the eigenvalues of the trained connectivity matrix. The theoretical predictions are indicated by the black dashed lines. Dots show results averaged over 200 network realizations. The shaded areas indicate the typical size of finite-size fluctuations, measured as the standard deviation of the collected sample.
		The target is fixed to $A=0.4$, parameters as in Fig.~\ref{fig:RLS_parallel}.
	}
	
	\label{fig:RLS_eig}
	
\end{figure}

As a final step, we look at the eigenvalues of the effective connectivity matrix $\chi + \mathbf{mn}^T$ (Eq.~\ref{eq:full_dynamics_rankone}).
The theory predicts that, for a solution in the form of Eq.~\ref{eq:sol}, one real outlier appears in the eigenspectrum. In a large network, its value corresponds to the overlap between the fedback vectors $\mathbf{m}^T\mathbf{n}$ \citep{Tao2013, MastrogiuseppeOstojic2}. 
We find that when the initial guess for the readout $\mathbf{n}$ is small, the amplitude of those additional components is weak (Fig.~\ref{fig:RLS_parallel}), and the eigenspectrum is close to the theoretical prediction (Fig.~\ref{fig:RLS_eig} {\bf a}, top). Spefically, the eigenspectrum includes a single real outlier whose position is well approximated by the scalar product  $\mathbf{m}^T\mathbf{n}$. The outlier eigenvalue gets closer to the predicted value as the training parameter $r$, which penalizes large readout weights, is increased (Fig.~\ref{fig:RLS_eig} {\bf b} and {\bf c}).
When the initial guess is characterized by strong scaling, instead, the shape of the eigenspectrum becomes very variable from one trial to the other. Typically, the eigenspectrum includes more than one outlier, the position of which fluctuates strongly independently of the theoretical prediction (Fig.~\ref{fig:RLS_eig} {\bf a}, bottom).

Taken together, those results suggest that the connectivity eigenspectrum may bear little information about the overall dynamics of the trained network. The eigenspectrum of the effective connectivity matrix $\chi + \mathbf{mn}^T$, indeed, appears to be very sensitive to the readout components which are spuriously introduced by the training algorithm but contribute little to the final output states. Our mean-field description, which specifically ignores such components, successfully captures the fundamental traits of local and global dynamics in trained networks but fails to predict the shape of the complex eigenspectra displayed by the synaptic connectivity matrices.

\subsection{Arbitrary geometry}

We complete our analysis by considering the RLS solutions which emerge from training in feedback architectures characterized by arbitrary initial geometries ($\rho<1$).

In Section 4, we found that the least-squares solution generates a region of local instability when the overlap between the two input vectors $\mathbf{m}$ and $\mathbf{I}$ is larger than a finite limit value (Fig.~\ref{fig:LS}). On the other hand, our theoretical framework suggests that -- as long as the two input vectors are not completely aligned ($\rho<1$) -- a stable mean-field solution always exists. The optimal readout $\mathbf{n}$ includes a strong component along $\eta_{I}$, i.e.~the direction of the external input $\mathbf{I}$ which is not shared with the feedback input vector $\mathbf{m}$  (Fig.~\ref{fig:general} {\bf d}).

\begin{figure}
	
	\centering
	\includegraphics{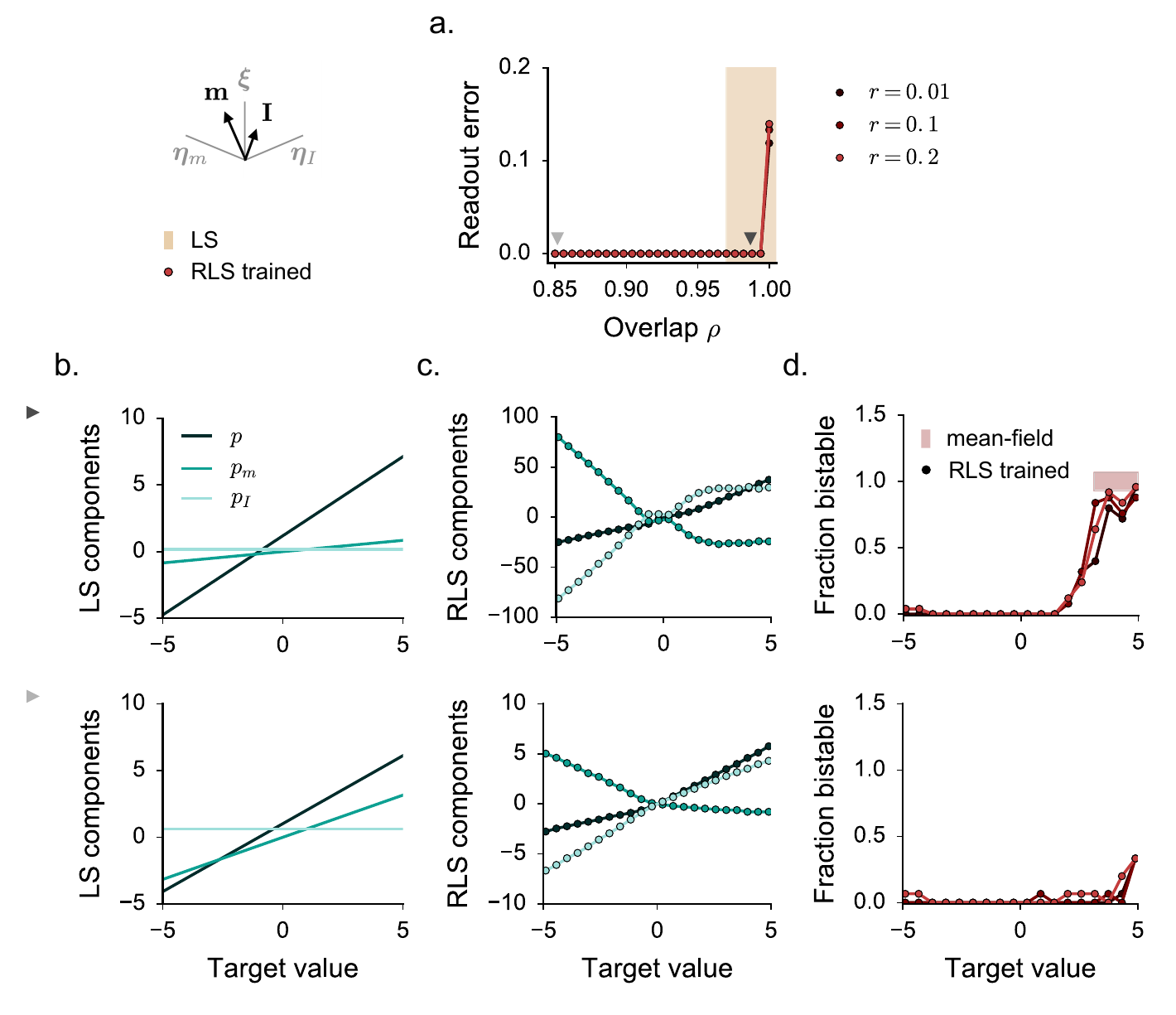}
	
	\caption{ 
		Recursive least-squares solutions (arbitrary $\rho$):  comparison between  mean-field predictions and trained networks.
		{\bf a.} Local stability of the target fixed point in feedback networks trained through the RLS algorithm. Details are as in Fig.~\ref{fig:RLS_parallel2} {\bf a}. We change the initial geometry of the network architecture by increasing values of the overlap $\rho$. The target value is fixed to $A=-1.1$. The yellow shaded area indicates the parameter region where the LS solution is unstable.
		{\bf b.} Geometry of the LS solution, characterized in terms of the analytical projections of the readout $\mathbf{n}$ onto the three axes $\bm{\xi}$, $\bm{\eta}_m$ and $\bm{\eta_I}$ (Eq.~\ref{eq:LS_weights}). The top and the bottom rows display results for two different values of $\rho$, indicated in {\bf a} by the black arrows.
		{\bf c.} Geometry of the RLS solution, computed numerically by projecting the readout $\mathbf{n}$ onto the three axes  $\bm{\xi}$, $\bm{\eta}_m$ and $\bm{\eta_I}$. 
		{\bf d.} Fraction of bistable trained networks, compared with the mean-field prediction extracted from the readout weights in {\bf c}. Details are as in Fig.~\ref{fig:RLS_parallel2} {\bf d}. Results are displayed for three different values of the regularization parameter $r$. Dots show results averaged over 15 network realizations.
		Parameters as in Fig.~\ref{fig:RLS_parallel}.
	}
	\label{fig:RLS_arbitrary}
	
\end{figure}

In Fig.~\ref{fig:RLS_arbitrary} {\bf a}, we select a value of the target for which the LS solution is unstable within a finite parameter region at large $\rho$ values. We then test the stability of the fixed point generated by FORCE training by continuously varying the value of the overlap parameter $\rho$.
In agreement with the mean-field theory, we find that the RLS algorithm systematically converges to a locally stable solution whenever the value of the overlap between the two input vectors is significantly smaller than one. 
Specifically, the algorithm converges inside and outside the shaded region where the least-square solution is unstable.

In order to understand how the instability is overcome, we fix the value of the overlap $\rho$ within the instability window, and we systematically compare the geometry of the LS and the RLS solutions (Fig.~\ref{fig:RLS_arbitrary} {\bf b}, {\bf c} and {\bf d} top).
As in Section 4, we extract the dominant readout geometries by projecting the readout vectors onto the three orthogonal axes  $\bm{\xi}$, $\bm{\eta}_m$ and $\bm{\eta_I}$  (Eq.~\ref{eq:sol}), resulting into sets of weights $(p, p_m, p_I)$ from which dynamics can be predicted.

We find that the LS solution contains strong components along $\bm{\xi}$ and $\bm{\eta}_m$, together with a weak and positive component along the non-shared input direction $\bm{\eta}_I$ (Fig.~\ref{fig:RLS_arbitrary} {\bf b} top). Importantly, because of Eq.~\ref{eq:LS_weights}, the value of $p_I$ is constant with respect to the target value $A$.
This solution generates a window of local instability at small target values (Fig.~\ref{fig:LS} {\bf b}) because of the interaction between the strong components along $\bm{\eta}_m$ and $\bm{\xi}$, which tend to produce symmetric and bistable solutions, and the small component along $\bm{\eta}_I$, which weakly suppresses bistability and disrupts such a symmetry. In particular, the instability area appears at negative target when the value of  $p_I$ is positive.

The RLS algorithm, on the other hand, converges to a readout vector which includes a strong component along $\bm{\eta}_I$ (Fig.~\ref{fig:RLS_arbitrary} {\bf c} top).
The value of $p_I$, furthermore, is not constant but varies with the value of the target $A$. In particular, the sign of $p_I$ coincides with the sign of the target. The instability region thus gets pushed to positive target values when $A$ is negative, and vice versa. 
The approximate mean-field description we derive from the RLS weights (Fig.~\ref{fig:RLS_arbitrary} {\bf c} top) confirms that such a region never emerges. Consistently, training always results in a locally stable fixed point.
 The mean-field description further predicts the existence of a second, locally stable solution at large and positive target values (Fig.~\ref{fig:RLS_arbitrary} {\bf d} top). In agreement, we find that a large fraction of trained networks displays bistability when the target $A$ is taken within such a window.

As a last step, we examine a network architecture characterized by a smaller value of the overlap $\rho$, taken outside of the LS instability window (Fig.~\ref{fig:RLS_arbitrary} {\bf b}, {\bf c} and {\bf d} bottom). We observe that, for smaller values of $\rho$, the bistability region predicted by the mean-field theory at large target values disappears  (Fig.~\ref{fig:RLS_arbitrary} {\bf d} bottom). Bistability is suppressed because, in correspondence of the same target values, the readout component along the non-shared input direction $\bm{\eta}_I$ is characterized by large relative weight (Fig.~\ref{fig:RLS_arbitrary} {\bf c} bottom). Consistently with the theory, the fraction of trained networks displaying bistability at large $A$ values significantly decreases (Fig.~\ref{fig:RLS_arbitrary} {\bf d} bottom).

To conclude, we tested the RLS algorithm on feedback networks characterized by arbitrary geometry, where our simplified theoretical framework predicts the existence of globally stable solutions. We found that the algorithm is systematically able to find readout solutions which generate the stable dynamics that is suitable to solve the task. Stability is achieved thanks to the flexibility of the online algorithm, which reshapes the geometry of the readout vector until a final optimal configuration is found. In contrast with the LS solution, such an optimal readout includes a strong and structured component along the non-shared input direction $\bm{\eta}_I$, which suppresses bistability and prevents local instabilities.

\section{Conclusions}

Predicting activity in trained recurrent networks is a challenging theoretical problem, due to the disordered synaptic structure and the large number of units \citep{Doya, Barak2017}. Through the recurrent dynamics, global bifurcations can appear as a result of minimal modifications to the synaptic weights and lead to dramatic changes in the network dynamical landscape. Among all the possible architectures, feedback networks represent a class of particularly simple recurrent networks. In these networks, training is restricted to a set of $N$ synaptic weights out of $N^2$, and the task-specific part of the recurrent dynamics is specified by the readout signal alone.

In this work, we examined how feedback networks can be trained to perform a single, stationary input-output association. Such a task is simple enough to be implemented in a purely feedforward setup, where no recurrent feedback exists and the reservoir dynamics is driven solely by the external input pattern. In such a case, which corresponds to setting $\sigma_m=0$ in Eq.~\ref{eq:mI}, the mean-field theory suggests that a task implementation can be constructed by aligning the readout vector $\mathbf{n}$ with the external input $\mathbf{I}$. This simple feedforward solution implements the task through a single stable fixed point, and thus possesses optimal stability properties.

Throughout the paper, we investigated how more complex solutions, which involve a rank-one recurrent feedback, can be constructed.
To this end, we derived a approximate mean-field framework which captures the geometry of the $N$-dimensional readout vector $\mathbf{n}$ with respect to the input vectors $\mathbf{m}$ and $\mathbf{I}$. 
In this theoretical setup, the random bulk connections are treated as quenched noise, so that correlations between the readout vector and the specific instatiation of the random matrix $\bm{\chi}$ are neglected. Note that the amplitude of the bulk connections $g$ affect the solutions of our mean-field equations only through minor quantitative differences. As a consequence, the space of solutions that our theory describes corresponds to network models where the global connectivity matrix is in practice rank-one  \citep{MastrogiuseppeOstojic2}.
We used the mean-field theory to show that implementing such a simple fixed-point task in a feedback network can generate output states characterized by non-trivial stability properties.
Specifically, we showed that the task can be solved if the readout $\mathbf{n}$ picks and amplifies the feedback component of the input, which is parallel to the vector $\mathbf{m}$ (Fig.~\ref{fig:general} {\bf c}). Similarly to \citet{Hopfield}, this implementation strongly relies on the non-linearity of the dynamics and produces additional spurious attractors. Alternatively, the task can be solved by aligning the readout vector with the external input $\mathbf{I}$ (Fig.~\ref{fig:general} {\bf d}). Such a configuration induces a hidden feedforward structure in the feedback network, which generates a single and stable output fixed point. In \emph{Appendix C}, we showed that this class of solutions performs optimally also in feedback networks characterized by a threshold-linear activation function.

The mean-field framework was further used to approximate the more complex, disordered solutions that are obtained via training with standard algorithmic techniques. 
Specifically, we used the theory to predict the local and the global stability of the target fixed point, and thus to understand how network parameters and target values impact training performance.
Despite implying very strong assumptions, we found that the mean-field approximation correctly describes trained networks in a surprisingly broad range of parameters. The approximation only fails to predict the value of the spurious output states in LS training when the architecture is strongly degenerate ($\rho \sim 1$) and the value of the random strength $g$ is large.
This non-trivial result indicates that the dynamical mechanisms underlying the input-output association in LS and RLS trained networks mimics the way the same computation emerge in a simple rank-one network \citep{MastrogiuseppeOstojic2}.

As a result of our theoretical analysis, we found that a common training algorithm, which relies on the least-square inversion of the constraint $\phi(\mathbf{x})^T \mathbf{n} = A$, returns under certain conditions suboptimal readout solutions. This happens because such a constraint, exactly like our system of mean-field equations (Eq.~\ref{eq:mf}), possibly admits multiple solutions but is blind with respect to the underlying dynamics which control stability. 
Furthermore, we found that both LS and RLS approaches systematically fail to converge to a locally stable solution when the geometry of the initial network architecture is degenerate, and the two input vectors $\mathbf{m}$ and $\mathbf{I}$ are parallel. Such a failure is explained by the approximate mean-field theory, which suggests that the unique task solution in the form of Eq.~\ref{eq:sol} is characterized by locally unstable dynamics. The mean-field framework, moreover, suggests that  introducing extra directions in the network geometry through the non-shared axes $\bm{\eta}_m$ and $\bm{\eta}_I$ expands the space of possible readout solutions. In Fig.~\ref{fig:RLS_arbitrary}, we showed that an online training algorithm like FORCE can take advantage of such a broader set of solutions by selecting the appropriate readout vector which generates stable output states.
Degenerate architectures consisting of strongly overlapping input vectors are not common in practical applications, where vectors $\mathbf{m}$ and $\mathbf{I}$ are typically generated at random. Note, however, that strong overlaps can be accidentally introduced in the network architecture when the two vectors $\mathbf{m}$ and $\mathbf{I}$ are uniform or generated through random distributions of large mean. In such cases, the overlap direction coincides with the unitary vector $\mathbf{u}$.

Our analysis indicates a possible important role for the parameter $g$, which scales the amplitude of the random bulk of connectivity. 
We found that increasing the value of $g$ does not significantly influence the structure of the feedback instabilities, which are mostly controlled by the geometry of the input vectors. Furthermore, large $g$ values have essentially no effect on training performance when the target fixed point can be implemented through uncorrelated, mean-field solutions. On the other hand, large random bulks significantly help the RLS algorithm to converge to strong, correlated solutions when uncorrelated solutions cannot satisfy the task. 
Correlated solutions are likely to play a major role in feedback networks trained to more complex tasks beyond the single input-output association. Understanding the mechanism through which large $g$ values actively supports training is thus an open but crucial question.	
As a hint, classical studies indicate that convergence of the RLS algorithm is guaranteed under the condition of \emph{persistent excitation} \citep{Bittanti, Kubin}. This condition requires that activity in the bulk, which serves as a set of basis functions, samples during training a large portion of the $N$-dimensional space spanned by the network population. From this perspective, large random bulks contribute to improve \emph{excitation} in two different directions: they slow down the decay of network activity, thus enlarging the training sample, and they increase the amplitude of the activity components that are orthogonal to the input vectors, thus increasing the sampling variance.

Our simplified mean-field description is complementary to the approach proposed in \cite{RivkindBarak}, where a similar fixed point task was studied. The mean-field analysis developed in \cite{RivkindBarak} is exact, and takes into account both the non-linear components of the readout vector and its correlations with the random bulk. Such an approach requires an analytical expression for trained readouts, and returns the local stability of the target fixed point. On the other hand, our analysis is approximate, but extends more easily to any readout vector obtained through algorithmic training. Our approach, furthermore, captures dynamics on a global scale, and allows us to predict the existence of spurious, bistable fixed points.

The mean-field framework we adopt is flexible enough to be directly extended to network architectures where the input weights are independently generated from non-Gaussian probability distributions or the activation function $\phi(x)$ has a different shape (see \emph{Appendix C}).
Significant additional work will be required, instead, to extend our results to networks of spiking units, for which different training algorithms have been proposed \citep{EliasmithAnderson, Boerlin2013, Thalmeier2016, Kim2018}.

The mean-field approach we considered here is directly adapted from \citet{MastrogiuseppeOstojic2}. Although the two studies build on the same theoretical tools, they are motivated by deeply different perspectives. In \citet{MastrogiuseppeOstojic2}, the low-rank part of the network connectivity is designed by hand: the overall understanding of the dynamics which emerges from the mean-field framework is exploited to construct low-rank connectivity structures which stably and efficiently implement a variety of behavioral tasks. Crucially, the tasks are in that case defined as qualitative input-output associations rules which fix the overall network behaviour but not the exact value of the readout $z$.
In this work, in contrast, the mean-field framework serves as a tool for understanding how an extremely simple task is implemented through training in standard network architectures. As in supervised learning applications, the task is specified by a fixed quantitative constraint on the readout value $z$. Consequently, the resulting target fixed point is not guaranteed to be a stable state for the network dynamics.
In \citet{MastrogiuseppeOstojic2}, moreover, both connectivity vectors $\mathbf{m}$ and $\mathbf{n}$ are considered to be plastic. In the network models that are derived, the two vectors serve two different computational roles: vector $\mathbf{n}$ selectively amplifies the external input patterns, while vector $\mathbf{m}$ defines the output direction to be picked by the readout. In the present framework, instead, the network architecture is constrained as in machine learning applications \citep{Jaeger, Lukosevicius, SussilloBarak}: the input weights $\mathbf{m}$ are generated at random and considered to be fixed, while the readout direction is given by the trained vector $\mathbf{n}$.

The mean-field analysis adopted in both studies is exact when two specific assumptions are satisfied: the readout vector $\mathbf{n}$ obeys a weak ($1/N$) scaling with the network size, and its entries are statistically uniform and uncorrelated with the random part of the connectivity $\bm{\chi}$. If these hypotheses hold, the resulting feedback network can only display simple or bistable stationary dynamics \citep{MastrogiuseppeOstojic2}, which was suitable in our case for implementing a simple fixed-point task.
For a feedback architecture trained to solve a more articulate task, on the other hand, the entries of  $\mathbf{n}$ need to be structured and/or fine-tuned to the random bulk. In that case, a mean-field description with the same characteristics would fail to correctly describe the network dynamics. However, one can hypothesize that -- as long as the task is reasonably simple -- the overall connectivity can still be approximated by the sum of a random noise term and a weak and uncorrelated connectivity matrix, where the rank of the latter is larger than one but still much smaller than the network size $N$. If such a low-dimensional decomposition can be extracted from the trained synaptic matrix, then a similar mean-field approach can be directly applied. Specifically, one can hope to predict dynamics by looking at the relative geometrical arrangement of a restricted number of high-dimensional vectors, which define the external inputs together with the low-rank part of the connectivity. 
Ongoing work indicates that this approach can be used to successfully extract the relevant dynamical mechanisms from recurrent networks trained on more complex tasks, where the spectrum of possible  network implementations is not known a priori.

\section*{Acknowledgements}
This work was funded by the Programme Emergences of City of Paris, Agence Nationale de la Recherche grants ANR-16-CE37-0016-01 and ANR-17-ERC2-0005-01, and the program ``Investissements d'Avenir'' launched by the French Government and implemented by the ANR, with the references ANR-10-LABX-0087 IEC and ANR-11-IDEX-0001-02 PSL University. The funders had no role in study design, data collection and analysis, decision to publish, or preparation of the manuscript.

\vspace{1cm}

\section*{Appendix A}

In this section, we provide details of the stability analysis for the stationary mean-field solutions, which correspond to the fixed points of the original dynamics (Eq.~\ref{eq:full_dynamics}). For every fixed point $\mathbf{x}$, stability is evaluated by computing the linear stability matrix and predicting the position of its eigenvalues.

The linear stability matrix reads: 
\begin{equation}
S_{ij} = \left( g\chi_{ij} + \frac{m_in_j}{N} \right) \phi'(x_j).
\end{equation}
As shown in Fig.~\ref{fig:general} {\bf b}, the eigenspectrum of $\mathbf{S}$ consists of two distinct components: a circular compact set of eigenvalues, directly inherited from the random bulk $\bm{\chi}$, and an isolated real outlier, mostly controlled by the rank-one feedback matrix $\mathbf{mn^T}$.
The radius of the circular set and the position of the outlier are evaluated by following the analysis in \citet{MastrogiuseppeOstojic2}.

The radius can be computed as in \citet{Rajan2006, Harish2015}:
\begin{equation}
r = g \sqrt{\langle [\phi'^2(x_i)]\rangle},
\end{equation}
where the average is evaluated as an integral over a Gaussian distribution of variance $\Delta$ (Eq.~\ref{eq:gauss_int}). In this work, we focus on fixed point dynamics, so we fix network parameters and target values to ensure $r<1$.

When the real outlier is not absorbed within the circular set, its position can be evaluated by computing the linearized dynamics of the average activation $\mu_i=[x_i]$ around the stationary mean-field solutions, corresponding to network fixed points. We illustrate here the main steps of the calculation; a more detailed account can be found in \cite{MastrogiuseppeOstojic2}.

In Eq.~\ref{eq:mf_i}, we derived that at the fixed point (that we here indicate by the apex $^0$) the average activation is given by: $\mu_i^0 = m_i z^0 + I_i$. After a small perturbation, we can write: $\mu_i(t) = \mu_i^0 + \mu_i^1(t)$, where the dynamics of $\mu^1(t)$ is given by:
\begin{equation}\label{eq:stab_mu1}
\dot{\mu}_i^1(t) = -\mu_i^1(t) + m_i z^1(t).
\end{equation}
Note that one can formally write: $\mu_i^1=m_i \tilde{z}^1$, where $\tilde{z}^1$ is the low-pass filtered version of $z^1$: $(1+\diff/\diff t) \tilde{z}^1 = z^1$.
Eq.~\ref{eq:stab_mu1} indicates that the decay time scale of the mean activity is inherited by the decay time constant of $z^1$. 
An additional equation for the time evolution of $z^1$ thus needs to be derived. 

When the network is perturbed, the firing activity $\phi_i=\phi(x_i)$ can be evaluated at the first order: $\phi_i^0 \rightarrow \phi_i^0 +\phi_i^1(t) = \phi(x_i^0) +\phi'(x_i^0)x_i^1(t)$. As a consequence, the first-order in $z$ reads:
\begin{equation}\label{eq:kappa1}
z^1(t) = \langle \bar{n}_i [\phi'(x_i^0)x_i^1(t)]\rangle.
\end{equation}
where $\bar{n}_i= N n_i$ and $n_i$ is as in Eq.~\ref{eq:sol}.
Summing Eq.~\ref{eq:kappa1} to its time-derivative, we get:
\begin{equation}
\dot{z}^1(t)=-z^1(t)+(1+\frac{\diff}{\diff t}) \langle \bar{n}_i [\phi'(x_i^0)x_i^1(t)]\rangle.
\end{equation}

As shown in \cite{MastrogiuseppeOstojic2}, one can first compute the average over the random connectivity, yielding:
\begin{equation}
\begin{split}
	[\phi'(x_i^0)x_i^1] & = m_i \ \tilde{\kappa}^1 [\phi_i'] + \left(  \frac{\Delta_0^1}{2} - \{ \langle \mu_i^1\mu_i^0 \rangle - \langle \mu_i^1\rangle \langle\mu_i^0 \rangle \} \right)[\phi_i''] \\
	& = m_i \ \tilde{\kappa}^1 [\phi_i'] + \left(  \frac{\Delta_0^1}{2} - \sigma_m^2 \tilde{z}^1z^0 - \sigma_{mI} \tilde{z}^1 \rangle \} \right)[\phi_i''],
\end{split}
\end{equation}
where the Gaussian integrals $[\phi_i']$ and  $[\phi_i'']$ are evaluated using the fixed point statistics.

As a second step, the average over units is performed, yielding a result in the form:
\begin{equation}
\langle \bar{n}_i [\phi'(x_i^0)x_i^1(t)]\rangle=  \tilde{z}^1 a + \Delta_0^1 \: b.
\end{equation}
A little algebra returns the value of the coefficients $a$ and $b$:
\begin{equation}\label{eq:ab}
\begin{split}
& a = c(p  \sigma_m \rho + p_m\sigma_m\sqrt{1-\rho^2})  \langle [\phi'(x_i)] \rangle \\
& b = \frac{c}{2} \left(p  \sigma_m \rho z^0 +  p_m\sigma_m \sqrt{1-\rho^2}z^0 + p  \sigma_I \rho+ p_I \sigma_I\sqrt{1-\rho^2} \right) \langle [\phi'''(x_i)] \rangle.
\end{split}
\end{equation}

The time evolution of $z^1$ can be finally rewritten as:
\begin{equation}\label{eq:kappa_ev}
\begin{split}
\dot{z}^1 (t) = - z^1 (t) + \left\{ a z^1  + b \left(1+\frac{\diff}{\diff t}\right) \Delta_0^1 \right\},
\end{split}
\end{equation}
so that the time evolution of the perturbed variance must be considered as well. By using Eq.~\ref{eq:full_dynamics}, one can easily show that, close to the fixed point:
\begin{equation}
\dot{\Delta}_0^1 = - \Delta_0^1 + \left\{ \mu^1 \frac{ \partial G }{\partial \mu}\Bigr|_{0} + \Delta_0^1 \frac{ \partial  G }{\partial \Delta_0}\Bigr|_{0}+z^1 \frac{ \partial  G }{\partial z}\Bigr|_{0} \right\}
\end{equation}
where $G(\mu, z, \Delta_0) = g^2 \langle [\phi_i^2(t)]\rangle + \sigma_m^2 z^2 + 2 \sigma_{mI} z + \sigma_I^2$. Explicitely computing the derivatives gives:
\begin{equation}
\begin{split}
&\frac{\partial G}{\partial \mu}\Bigr|_0  =2 g^2 \langle [\phi_i\phi_i']\rangle \\
&\frac{\partial G}{\partial \Delta_0}\Bigr|_0= g^2\left\{\langle [\phi_i'^2]\rangle+ \langle [\phi_i\phi_i'']\rangle \right\}\\
&\frac{\partial G}{\partial z}\Bigr|_0= 2\sigma_m^2z^0 + 2 \sigma_{mI}.
\end{split}
\end{equation}
As the input vectors and the bulk connectivity have zero mean, furthermore: $\dot{\mu^1}(t)=-\mu^1(t)$.

We finally obtained that the perturbation time scale is determined by the population-averaged dynamics:
\begin{equation}\label{eq:kappa_dyn}
\frac{\diff}{\diff t}\begin{pmatrix}  \mu^1 \\ \Delta_0^1 \\ \kappa^1\end{pmatrix} = - \begin{pmatrix} \mu^1 \\ \Delta_0^1 \\ \kappa^1 \end{pmatrix} + \mathcal{M} \begin{pmatrix} \mu^1 \\ \Delta_0^1 \\ \kappa^1 \end{pmatrix} 
\end{equation}
where the evolution matrix $\mathcal{M}$ is defined as:
\begin{equation}\label{eq:M}
\mathcal{M} = 
\begin{pmatrix} 0 & 0 & 0\\ 
2 g^2 \langle [\phi(x_i)\phi'(x_i)]\rangle & g^2\left\{\langle [\phi'(x_i)^2]\rangle + \langle [\phi(x_i)\phi''(x_i)]\rangle \right\} & 2\sigma_m^2z^0 + 2\sigma_{mI}\\
2 b g^2 \langle [\phi(x_i)\phi'(x_i)]\rangle & b g^2\left\{\langle [\phi'(x_i)^2]\rangle + \langle [\phi(x_i)\phi''(x_i)]\rangle \right\} & b  (2\sigma_m^2z^0 + 2\sigma_{mI}) + a
\end{pmatrix}.
\end{equation}
When an outlier eigenvalue is present in the stability eigenspectrum, its position can be finally evaluated as the largest eigenvalue of the reduced stability matrix $\mathcal{M}$.

\section*{Appendix B}

The study in \citet{RivkindBarak} addresses the local stability of the full least-squares solution. The position of the outlier eigenvalue $\lambda$ is in that case computed by combining a large network mean-field description with control theory arguments. 
Following the same calculations, we compute the value of $\lambda$ in the current and more general setup, yielding:
\begin{equation}
\begin{split}
\lambda &= \frac{g^2}{\Delta_i} \left\{ - \langle [\phi^2(x_i)] \rangle + \langle(m_i A + \sqrt{\Delta_i}w_i)[ \phi(x_i) \phi'(x_i)]\rangle \right\} + 1\\
& = g^2 \left\{ \langle [\phi'^2(x_i) ] \rangle + \langle[ \phi(x_i) \phi(x_i)'' ]\rangle \right\} \left(1 + \frac{A}{\Delta_i}\left[\sigma_m\rho^2 (\sigma_m A + \sigma_I ) + \sigma_m^2 (1-\rho^2) A \right]\right),
\end{split}
\end{equation}
where $\Delta_i = g^2 \langle[\phi^2(x_i)]\rangle$ as in Eq.~\ref{eq:mf_i}. In Fig.~\ref{fig:LS}, this expression is evaluated for the mean-field solution corresponding to the target (light green traces).

\section*{Appendix C}

In this section, we extend our results to feedback networks characterized by a threshold-linear activation function. We set: $\phi(x) = [x - T]_+$, where $T$ is the parameter controlling the activation threshold. Such a piecewise-linear activation function is extremely popular in machine learning applications based on feedforward networks \citep{Krizhevsky}.  Recent literature focusing on recurrent architectures \citep{Kadmon2015, MastrogiuseppeOstojic, RivkindBarak}, on the other hand, seems to indicate that threshold-linear functions (more than sigmoids) can impair dynamical stability in simple recurrent models. As an example, we focus on negative threshold values ($T=-0.5$), which seem to guarantee better stability for the fixed-point task we consider \citep{RivkindBarak}.

\begin{figure}[t]

	\centering
	\includegraphics{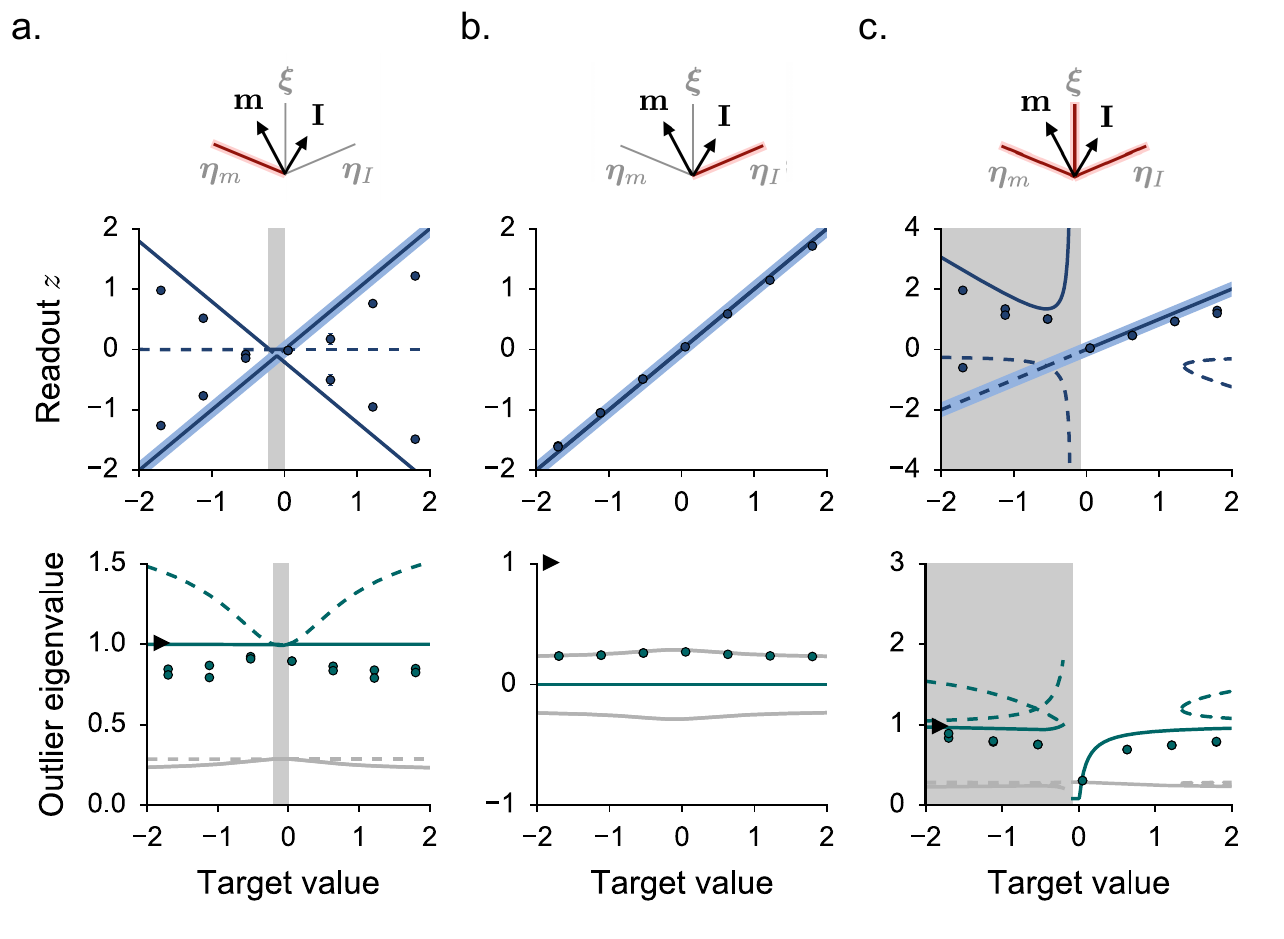}
	
	\caption{ 
		Implementing the fixed point task with a threshold-linear network ($\phi(x) = [x-T]_+$, $T=-0.5$): mean-field analysis in three example geometries.
		{\bfseries a-b-c.} Mean-field characterization of the network output states for three different readout geometries (see Fig.~\ref{fig:general}). Continuous (resp. dashed) lines correspond to locally stable (resp. unstable) mean-field solutions. Top row: value of the readout signal. Bottom row: position of the outlier eigenvalue in the stability eigenspectra. Details as in Fig.~\ref{fig:general} {\bfseries c-d-e}.
		The results of simulations are displayed as dots ($N=6000$, average over 20 network realizations). We integrate numerically the dynamics of finite-size networks where the readout vector $\mathbf{n}$ is normalized through Eq.~\ref{eq:c}. In panels {\bf a} and {\bf c}, we observe large finite-size effects due to the outlier stability eigenvalues laying very close to the instability boundary.
		Parameters as in Fig~\ref{fig:general}. 
	}
	\label{fig:general_relu}
	
\end{figure}

As a first step, as in Fig.~\ref{fig:general}, we select arbitrary values of $(p, p_m, p_I)$ (Eq.~\ref{eq:sol}) to construct example network implementations. We analyze these networks by solving the resulting mean-field equations to predict the output states and their stability.
The mean-field equations illustrated in Section 3 have been derived by assuming a generic activation function $\phi(x)$, so they directly generalize to threshold-linear networks. The Gaussian integrals in Eqs.~\ref{eq:mf} and \ref{eq:M} can in this case be evaluated analytically (note that integrals involving the high-order derivatives $\phi''$ and $\phi'''$ systematically vanish).
Results are displayed in Fig.~\ref{fig:general_relu}. 

Similarly to Fig.~\ref{fig:general} {\bf d}, Fig.~\ref{fig:general_relu} {\bf b} indicates that aligning the readout vector $\mathbf{n}$ with the private component of the external input vector $\bm{\eta_I}$ returns a unique and stable target fixed point. The position of the outlier eigenvalue in the stability eigenspectrum, which is controlled by $\phi'''$ (Eq.~\ref{eq:ab}), vanishes in the present case. We conclude that the optimal solution derived in the case of a sigmoidal activation function, which consists of a readout vector aligned with $\bm{\eta_I}$, applies to threshold-linear networks as well. 

Figs.~\ref{fig:general_relu} {\bf a} and {\bf c}, on the other hand, indicate that readout solutions which contain components along the feedback input vector $\mathbf{m}$ are characterized by poor stability properties. When the readout is aligned with the non-shared component of $\mathbf{m}$ (Fig.~\ref{fig:general_relu} {\bf a}), two marginally stable fixed points (characterized by an outlier eigenvalue laying exactly on the instability line) are generated.
This implementation of the task is characterized by extremely long time scales and large finite-size effects, and thus results in large readout errors.
When the readout contains components along both input vectors $\mathbf{m}$ and $\mathbf{I}$  (Fig.~\ref{fig:general_relu} {\bf c}), instead, the two fixed points behave asymmetrically.
The positive branch of the mean-field solution, which corresponds to the target fixed point when $A$ is positive, is formally stable, but is characterized by a very large outlier eigenvalue. The negative branch of the solution is instead systematically unstable, and generates a very broad instability window at negative target values (grey area in Fig.~\ref{fig:general_relu} {\bf c}). Both in Figs.~\ref{fig:general_relu} {\bf a} and {\bf c}, large outlier eigenvalues are given by the average of the third-order derivative $\langle [\phi_i'''] \rangle$ vanishing in Eqs.~\ref{eq:M} and \ref{eq:ab}. This term quantifies the effect of saturation and has a stabilizing effect in networks characterized by the sigmoidal activation function (Fig.~\ref{fig:general}). Note that, as already observed in a previous study \citep{MastrogiuseppeOstojic}, threshold-linear networks appear to be characterized by larger finite-size effects than sigmoidal ones.

\begin{figure}[t]
	\begin{adjustwidth}{-0.in}{0in} 
		\centering
		
		\includegraphics{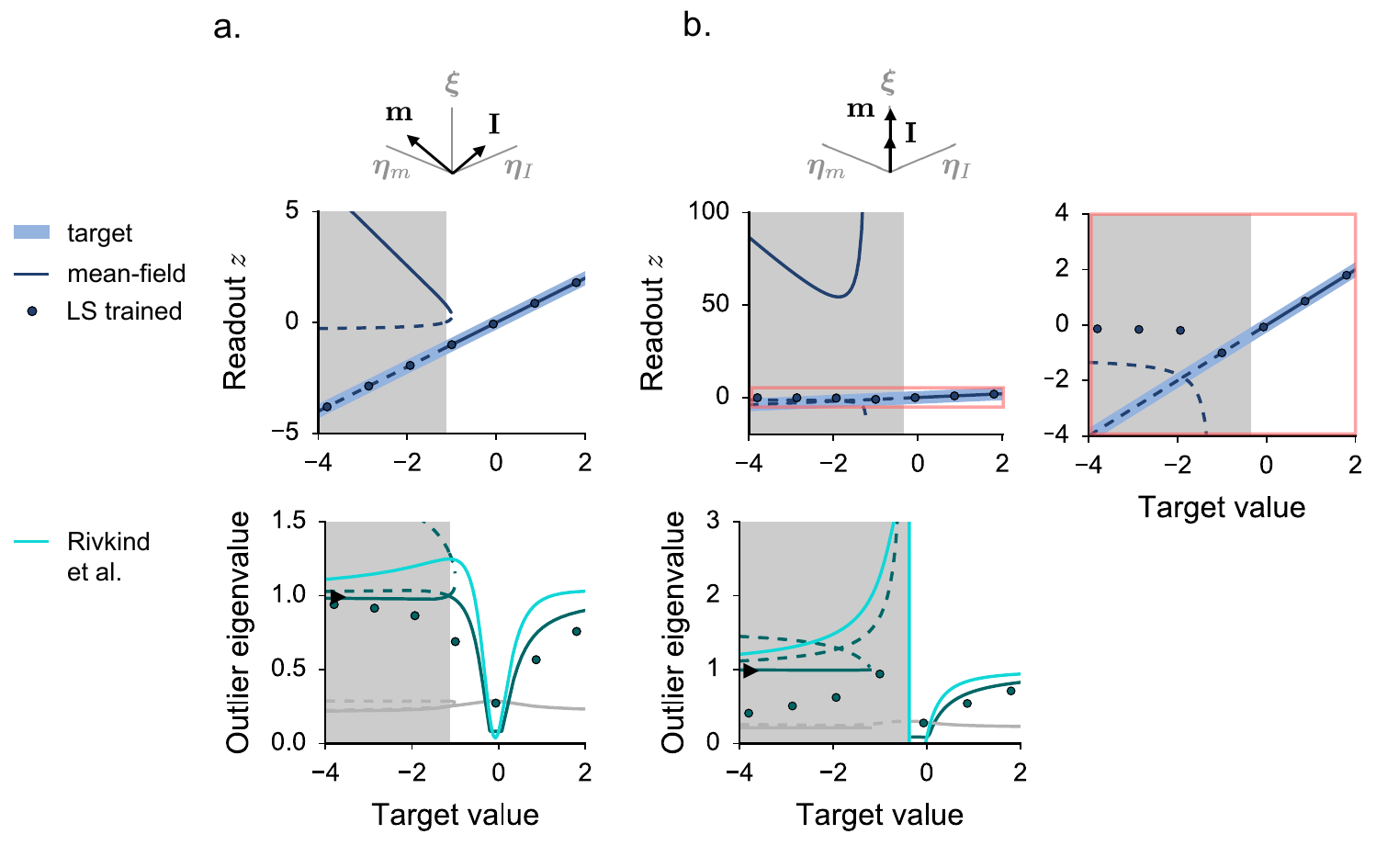}
	\end{adjustwidth}
	
	\caption{ 
		The least-squares solution in a threshold-linear network ($\phi(x) = [x-T]_+$, $T=-0.5$): approximate mean-field description. 
		{\bf a-b.} Mean-field characterization for three different configurations of the non-trained input vectors $\mathbf{m}$ and $\mathbf{I}$. In {\bf a}, we take $\rho=0.6$; in {\bf b}, $\rho=1$. Continuous and dashed lines indicate the solutions of the simplified mean-field description, where the readout is approximated by a vector in the form of Eq.~\ref{eq:sol}. Details are as in Fig.~\ref{fig:LS}. The light green line indicates the exact value of the outlier eigenvalue which measures the local stability of the target solution \citep{RivkindBarak} (details provided in \emph{Appendix B}). 
		The right panel in {\bf b} top is a magnified view of the area highlighted in pink in the left panel.
		Dots display the results of simulations from finite-size networks ($N=6000$, averages over 20 network realizations). We integrate numerically the dynamics of finite-size networks where the readout vector $\mathbf{n}$ is given by the LS solution (Eq.~\ref{eq:LS}). In order to reach different solutions, we initialize the network dynamics in two different initial conditions, centered around $\mathbf{n}$ and $-\mathbf{n}$. 
		Parameters as in Fig.~\ref{fig:general}. 
	}
	\label{fig:LS_relu}
	
\end{figure}

In a second step, similarly to Fig.~\ref{fig:LS}, we investigate how threshold-linear networks implement the fixed point task when they are trained via LS inversion (Fig.~\ref{fig:LS_relu}). 
We find that mean-field solutions are again characterized by large values of the outlier eigenvalue. When the value of the target is negative, specifically, the target solution is predicted to be unstable both from our approximate and the exact \citep{RivkindBarak} theory. This instability window further appears both in network architectures characterized by small and large values of the overlap $\rho$ (Fig.~\ref{fig:LS_relu} {\bf a} and {\bf b}).

In Fig.~\ref{fig:LS_relu}, we compare theoretical results with simulations performed in finite-size networks trained via LS inversion. In contrast to the prediction of the theory we find that, when the value of the overlap $\rho$ is small, finite-size networks manage to keep the readout error small. The output fixed point, however, displays extremely long timescales. When the value of $\rho$ is large, on the other hand, the readout value significantly diverges from the target when $A$ is negative. 

To conclude, we briefly investigated how threshold-linear networks solve the fixed point task. We found that the readout components aligned with the feedback vector $\mathbf{m}$ do not robustly generate bistability, but give rise to instabilities and dramatically slow down dynamics by pushing the value of the outlier eigenvalue close to the instability boundary. Finally, for such a class of networks, it could be interesting to further analyze readout solutions characterized by strongly non-linear geometries (see Section 4.2). We keep this direction open for further studies.

\section*{Appendix D}

In this section, we consider the least-squares readout $\mathbf{n}$ discussed in Section 4, and we develop a mean-field description which takes the effect of the non-linearity $\phi(x)$ into account (Fig.~\ref{fig:nonlinear}).
For the sake of simplicity, we consider the scenario where the input vectors $\mathbf{m}$ and $\mathbf{I}$ are parallel ($\rho=1$), so that the  $\bm{\xi}$ is the only dominant direction to be taken into account ($p_m=p_I=0$). Such setup is of specific interest, as it corresponds to the case where the mean-field approximation deviates more strongly from the results obtained through simulating dynamics in trained networks. The non-linear mean-field description we derive here indicates which fraction of such deviations can be explained by including non-linear components in our theoretical description.

The full LS solution is proportional to (Eq.~\ref{eq:LS}):
\begin{equation}
\phi(x_i^{ol}) =\phi\left( (\sigma_m A + \sigma_I) {\xi_i} + \delta_i^{ol} \right).
\end{equation}
We approximate $\bm{\delta}^{ol}$ by white noise of self-consistent amplitude, thus yielding an approximate readout in the form:
\begin{equation}
{{n}_i} = c \: \phi\left( (\sigma_m A + \sigma_I) {\xi_i} + \sqrt{\Delta_i^{ol}} w_i^{ol}  \right),
\end{equation}
where $w_i^{ol}$ is a standard Gaussian variable.
With this choice, the analytical expression for $z$ reads:
\begin{equation}
z = c \int \mathcal{D} \xi  \int \mathcal{D} w^{ol} \: \phi\left( (\sigma_m A + \sigma_I) \xi + \sqrt{\Delta_i^{ol}}w^{ol} \right) \int \mathcal{D} w \: \phi\left(  (\sigma_m z + \sigma_I) \xi + \sqrt{\Delta_i}w \right),
\end{equation}
which can be solved together with the usual expression for the second-order statistics (Eq.~\ref{eq:mf}).
Crucially, the residual input $\sqrt{\Delta_i}w$ is not constrained to match the residual input in the open-loop condition ($\sqrt{\Delta_i^{ol}}w^{ol}$), as additional output states -- together with the target fixed point -- can exist.

As in Eq.~\ref{eq:c}, we fix the normalization factor $c$ by matching the network output with the target in the open-loop configuration. We thus impose $z=A$, $\Delta_i=\Delta_i^{ol}$ and $w=w^{ol}$, that gives:
\begin{equation}
c = \frac{A}{ \int \mathcal{D} \xi \left[ \int \mathcal{D} w^{ol} \: \phi\left(  (\sigma_m A + \sigma_I) \xi + \sqrt{\Delta_i^{ol}}w^{ol} \right) \right]^2}.
\end{equation}

\section*{Appendix E}

In this section, we illustrate in detail the recursive least-squares (RLS) algorithm \citep{SussilloAbbott} used in Section 5.

The algorithm runs online; that is, weights modifications are alternated with short time windows where the network dynamics is simulated. Specifically, we update the readout weights every $\Delta t =$ 0.1 $\tau$, where $\tau$ is the integration time constant of a single unit, which is here normalized to one. The time step of integration is instead taken equal to 0.01 $\tau$. As the task is extremely simple, the total training time is taken equal to 150 $\tau$. 

At the beginning of training, the readout vector $\mathbf{n}$ is initialized as a Gaussian vector of std $N^{s}$ (see Figs.~\ref{fig:RLS_parallel} and \ref{fig:RLS_parallel2}). If not differently stated, we fix $s = - 0.5$. At every training step, the components of $\mathbf{n}$ are updated by following \citet{SussilloAbbott}. 
To begin with, the error $e(t) = z(t) - A$ is computed. Then, the readout vector $\mathbf{n}$ is updated according to:
\begin{equation}
\mathbf{n}(t) = \mathbf{n}(t - \Delta t) - e(t) \: \mathbf{P} (t) \: \phi (\mathbf{x}(t)).
\end{equation}
The matrix $\mathbf{P}$ is a running extimate of the inverse of the rate correlation which includes a regularization term \citep{SussilloAbbott}. At the beginning of training, $\mathbf{P}$ is initialized as:
\begin{equation}
\mathbf{P}(0) = \frac{\mathbf{I}}{r}
\end{equation}
where $\mathbf{I}$ is the identity matrix and $r$ is a scalar that we refer to as the regularization parameter. If not differently stated, the value of $r$ is fixed to 0.1. At every training step, $\mathbf{P}$ is updated as:
\begin{equation}
\mathbf{P} (t) = \mathbf{P} (t-\Delta t) - \frac{ \mathbf{P} (t-\Delta t) \: \phi (\mathbf{x}(t)) \: \phi (\mathbf{x}(t))^T \: \mathbf{P} (t-\Delta t) }{1+  \phi (\mathbf{x}(t))^T \: \mathbf{P} (t-\Delta t) \: \phi (\mathbf{x}(t))}.
\end{equation}

\end{document}